\documentclass[10pt,journal,compsoc]{IEEEtran}
\ifCLASSOPTIONcompsoc
  \usepackage[nocompress]{cite}
\else
  \usepackage{cite}
\fi
\ifCLASSINFOpdf
  \usepackage[pdftex]{graphicx}
  \graphicspath{{../pdf/}{../jpeg/}}
  \DeclareGraphicsExtensions{.pdf,.jpeg,.png}
\else
  \usepackage[dvips]{graphicx}
  \graphicspath{{../eps/}}
  \DeclareGraphicsExtensions{.eps}
\fi
\usepackage{amsmath}
\usepackage{algorithmic}

\usepackage{array}

\ifCLASSOPTIONcompsoc
  \usepackage[caption=false,font=footnotesize,labelfont=sf,textfont=sf]{subfig}
\else
  \usepackage[caption=false,font=footnotesize]{subfig}
\fi
\usepackage{fixltx2e}

\ifCLASSOPTIONcaptionsoff
  \usepackage[nomarkers]{endfloat}
 \let\MYoriglatexcaption\caption
 \renewcommand{\caption}[2][\relax]{\MYoriglatexcaption[#2]{#2}}
\fi
\usepackage{url}

\usepackage[shortlabels]{enumitem}
\usepackage{booktabs}
\usepackage{listings}
\usepackage{fourier} 
\usepackage{array}
\usepackage{makecell}
\usepackage{rotating}
\usepackage{pdflscape}
\usepackage{afterpage}
\usepackage{capt-of}
\usepackage{lscape}
\usepackage{xcolor}
\usepackage{xspace}
\usepackage{color, colortbl}
\usepackage{graphicx}

\definecolor{Gray}{gray}{0.9}
\definecolor{LightCyan}{rgb}{0.88,1,1}
\definecolor{LightGray}{rgb}{0.82,0.82,0.82}
\definecolor{BrightGray}{rgb}{0.92,0.92,0.92}

\usepackage{bookmark}

\begin{document}

\newcommand{\refactor}[1]{\textcolor{red}{\textbf{[#1]}}}
\newcommand{\fix}[1]{\textcolor{red}{#1}}
\newcommand{\added}[1]{\textcolor{blue}{#1}}
\newcommand{\modified}[1]{\textcolor{purple}{#1}}

\newcommand{\cooltitle}[1]{\vspace{1mm}\noindent\textbf{#1}}
\newcommand{\cooltitlecomb}[1]{\vspace{1mm}\textbf{\textit{#1}}}

\newcommand{\numberofdrivers}{33\xspace}
\newcommand{\numberofinterviews}{28\xspace}
\newcommand{\numberofengineers}{19\xspace}
\newcommand{\numberofsecuritypro}{9\xspace}
\newcommand{\numberofmemberchecks}{12\xspace}
\newcommand{\numberofnewdrivers}{8\xspace}
\newcommand{\frameworkname}{DASP\xspace}

\newcommand{\enrique}[1] {\textcolor{blue}{\textbf{[Enrique: #1]}}}
\newcommand{\omar}[1] {\textcolor{gray}{\textbf{[Omar: #1]}}}
\newcommand{\peggy}[1] {\textcolor{purple}{\textbf{[Peggy: #1]}}}

\newcommand{\idecode}[0]{non-notebook code}
\renewcommand{\theenumi}{\Alph{enumi}}

%
\title{\frameworkname: A Framework for Driving the Adoption of Software Security Practices}

%
%
%
%

\author{Enrique~Larios-Vargas,~\IEEEmembership{Member,~IEEE,}
        Omar Elazhary,~\IEEEmembership{}
        Soroush Yousefi,~\IEEEmembership{}
        Derek Lowlind,~\IEEEmembership{}
        Michael~L.~W.~Vliek,~\IEEEmembership{}
        and Margaret-Anne~Storey,~\IEEEmembership{Member,~IEEE}
\IEEEcompsocitemizethanks{\IEEEcompsocthanksitem E. Larios-Vargas, O. Elazhary, S. Yousefi, D. Lowlind, and M. A. Storey are with the Department
of Computer Science, University of Victoria, Canada.\protect\\
E-mail: elariosvargas@uvic.ca
\IEEEcompsocthanksitem M. Vliek is with Leiden University.}
\thanks{Manuscript received April 19, 2005; revised August 26, 2015.}}

%
%

\markboth{Journal of \LaTeX\ Class Files,~Vol.~14, No.~8, August~2015}%
{Shell \MakeLowercase{\textit{et al.}}: Bare Demo of IEEEtran.cls for Computer Society Journals}
%



\IEEEtitleabstractindextext{%
\begin{abstract}
Implementing software security practices is a critical concern in modern software development. 
Industry practitioners, security tool providers, and researchers have provided 
 standard security guidelines and sophisticated security development tools to ensure a secure software development pipeline. 
But despite these efforts, there continues to be an increase in the number of vulnerabilities that can be exploited by malicious hackers.
There is thus an urgent need to understand why developers still introduce security vulnerabilities into their applications and to understand what can be done to motivate them to write more secure code. 
%
To understand and address this problem further, we propose \frameworkname, a framework for diagnosing and driving the adoption of software security practices among developers. \frameworkname was conceived by combining behavioral science theories to shape a cross-sectional interview study with \numberofinterviews software practitioners. Our interviews lead to a framework that consists of a comprehensive set of \numberofdrivers drivers grouped into 7 higher-level categories that represent what needs to happen or change so that the adoption of software security practices occurs. Using the \frameworkname framework, organizations can design  interventions suitable for developers' specific development contexts that will motivate them to write more secure code.

\end{abstract}

}

\maketitle

\IEEEdisplaynontitleabstractindextext

%
\IEEEpeerreviewmaketitle

	\section{Introduction}
	\label{sec:introduction}

Software security is undeniably one of the most critical 
and ongoing concerns in modern software development. 
In 2021, the NIST Computer Security Division\footnote{https://nvd.nist.gov/} identified over 18,000 software vulnerabilities, 
and this number has been steadily increasing from 2016~\cite{nvd2021}. 
Flawed applications might behave unpredictably, and these weaknesses are often abused by malicious hackers~\cite{veracode2021}. 
For example, recent reports show that malicious attackers use unique platforms and search engines, e.g., Shodan\footnote{https://www.shodan.io/}, to scan for networks that are exposed to known vulnerabilities and exploit them before a victim can apply a patch~\cite{portswigger2021}.

Software developers are responsible for many of these vulnerabilities. 
Notably, around 60\% of 
vulnerabilities identified by Veracode\footnote{https://info.veracode.com/report-state-of-software-security-volume-11} in a study of 130,000 active 
applications highlighted that a developer's lack of careful development and 
maintenance was a significant reason for the introduction of vulnerabilities. 
These vulnerabilities occur because
developers face pressure to meet customer requirements and deliver features quickly.
In addition, they often treat security as a less important non-functional requirement, typically not considering security until the 
last stages of the software development life cycle. Often security is not a concern unless security compliance is imposed by employers or application users~\cite{assal2019}. 
The delayed consideration of security issues makes it  
more challenging and even more expensive to address in later stages~\cite{poller2017}.

Neglecting software security is a well-recognized problem in any industry. As a result, there are many ongoing efforts to fix it by leading cybersecurity organizations of different business types, such as the MITRE corporation\footnote{https://attack.mitre.org/}, OWASP\footnote{https://owasp.org/}, the CERT Division\footnote{https://www.sei.cmu.edu/about/divisions/cert/}, and NIST\footnote{https://www.nist.gov/}. These organizations provide security standards and excellent resources to help practitioners ensure a secure software product.
There are also hundreds of free online resources available for practitioners to learn software security practices. 
Furthermore, there is a huge active community of security professionals and researchers behind the development of security tools and keeping up-to-date security guidelines aimed at ensuring a secure software development life cycle.
However, despite the availability of these many resources, developers continue to introduce security vulnerabilities in source code, and organizations still lack proper guidelines for designing strategies to mitigate poor security. 

This situation pushes forward the need to properly understand developer behaviors---specifically, what drives developers to adopt software security practices. With this knowledge, organizations would be better positioned to design interventions to foster behavior change, leading developers to write more secure code. To understand developer behaviour towards the adoption of security practices, we conducted a cross-sectional interview study with a cohort of 28 software practitioners and used the COM-B Model for behavior change~\cite{michie2011} as a diagnosis tool to understand the capabilities, opportunities, and motivations behind the adoption of software security practices.

The insights from our study have led to a novel actionable framework that organizations and developers can use to drive the adoption of software security practices. The framework captures \numberofdrivers drivers across seven categories of behavior change. In addition, we exhaustively compared our findings with current literature, noticing that previous research did not report three of our drivers. Our framework can be used by organizations to diagnose security challenges and to design strategies that influence the adoption of security practices within their specific context. This is the first study that considers behavior change in software security, opening the door for future research. Our framework can support future researchers by leveraging the power of well-known behavioral psychology theories to understand and drive improvement in secure software development.

The following section (Section 2) provides some background on the behavioral theories that shaped the design of our study. Then we describe the cross-sectional interview methodology and our analysis approach in Section 3. We present our findings in Section 4,and review related work in Section 5. Next, we discuss how organizations can use our framework in Section 6 and provide implications for developers, security specialists, and researchers. In Section 7, we detail the threats to the validity of our research and conclude the paper in Section 8.

	\section{Background}
	\label{sec:background}

\begin{table*}[]
	\centering
	
	\caption{Intervention function definitions from  the BCW.}
    \resizebox{0.8\textwidth}{!}{%
	
	\begin{tabular}{ l  l }
		
		\toprule
	\textbf{Intervention function}	& \textbf{Definition}  \\
		
		\midrule
		
Education & Increasing knowledge or understanding   \\
Persuasion &  Using communication to induce positive or negative feelings or stimulate action \\
Incentivization & Creating an expectation of reward   \\
Coercion & Creating an expectation of punishment or cost   \\
Training & Imparting skills  \\
Restriction & \thead[l]{Using rules to reduce the opportunity to engage in the target behaviour (or to \\ increase the target behaviour by reducing the opportunity to engage in competing \\ behaviours)}  \\
Environmental restructuring & Changing the physical or social context   \\
Modeling & Providing an example for people to aspire to or imitate \\
Enablement &  \thead[l]{Increasing means/reducing barriers to increase capability (beyond education and \\ training) or opportunity (beyond environmental restructuring)} \\

		\bottomrule
		
	\end{tabular}%
}
	\vspace{0.02in}
	
	\label{tab:interventionfunction}
	\vspace{-2mm}
	
\end{table*}
\begin{table*}[]
	\centering
	
	\caption{Policy category definitions from  the BCW.}
    \resizebox{0.8\textwidth}{!}{%
	
	\begin{tabular}{ l  l }
		
		\toprule
	\textbf{Policy categories}	& \textbf{Definition}  \\
		
		\midrule
		
Communication/marketing & Using print, electronic, telephonic, or broadcast media   \\
Guidelines &  \thead[l]{Creating documents that recommend or mandate practice. This includes all \\
changes to service provision} \\
Fiscal measures & Using the tax system to reduce or increase the financial cost   \\
Regulation & Establishing rules or principles of behaviour or practice   \\
Legislation & Making or changing laws  \\
Environmental/social planning & Designing and/or controlling the physical or social environment \\
Service provision & Delivering a service  \\
		\bottomrule
		
	\end{tabular}%
}
	\vspace{0.02in}
	
	\label{tab:policycategory}
	\vspace{-5mm}
	
\end{table*}
\begin{table*}[]
	\centering
	
	\caption{COM-B model components and examples in the context of software security.}
	 \resizebox{0.9\textwidth}{!}{%
	
	\begin{tabular}{ l  l  l}
		
		\toprule
	\textbf{COM-B model component} & \textbf{Description}  & \textbf{Examples} \\
		
		\midrule
		
Technical capability & Having the technical skills to perform security practices & \thead[l]{Having the skill to understand technical aspects of security \\ exploits and apply security patches}  \\
Non-Technical capability & Having the non-technical skills to perform security practices & \thead[l]{Having the skill to communicate and discuss security issues \\ with all stakeholders compromised in a security incident}\\
Psychological capability &  \thead[l]{Being aware of the negative consequences of not adopting\\
security practices \\ Having the knowledge or confidence to apply security practices} &  \thead[l]{Awareness of the compliance standards that regulate\\ data protection and privacy, such as GDPR, PCI, and HIPPA}\\
Technical opportunity & \thead[l]{Having the opportunity afforded by the environment involving\\ time, resources, tools, and locations }  &  \thead[l]{Having tools that can be easily integrated into the \\ development workflow and alert developers of potential risks}\\
Social opportunity & \thead[l]{Having the opportunity afforded by interpersonal influences, \\ social and cultural norms that influence the way developers\\ 
	think about things}   & \thead[l]{Having people around implementing security practices\\
	  reminds developers why to invest extra effort in security} \\
Reflective motivation &  \thead[l]{Reflective processes involving plans (self-conscious intentions)\\
	and evaluations (beliefs about what is good and bad)} & \thead[l]{Intending to follow security guidelines after understanding \\their value and the rationale behind them} \\
Automatic motivation &  \thead[l]{Automatic processes involving emotional reactions, desires,\\
	(wants and needs), impulses, inhibitions, drive states, and\\reflex responses}  & \thead[l]{Developers feeling frustrated due to the lack of support from\\ management when prioritizing security over other tasks} \\

		\bottomrule
		
	\end{tabular}%
}
	\vspace{0.02in}
	
	\label{tab:combcomponents}
	\vspace{-2mm}
	
\end{table*}

Understanding human behavior is challenging due to the diverse number of psychological factors that may impact positive and negative change. As a result, behavioral science disciplines, such as psychology and behavioral economics, offer many theories and models that describe drivers of behavior, such as attitudes, motivations, norms, habits, and behavioral control. Many of these theories and models can be further used to predict human decision-making and behaviors (e.g.,\cite{hovland1953,fishbein1975,janz1984,kahneman1979,strack2004,bandura1982}). 
Similarly, it is a challenge to understand what drives change in human behaviors in software development activities such as software security, but doing so may help organizations understand how to improve the adoption of practices that will encourage developers to write more secure code. 
In the following, we describe three behavioral theories and models used to study behavior change. Later we discuss how we used these theories to design our novel study of the potential for behavior change in software security.

\subsection{Behavioral Theories and Models}
We present three behavioral theories and models: (1) the COM-B model and the Behavior Change Wheel, (2) the Self-Efficacy Theory, and (3) the Response efficacy concept. 

\subsubsection{The COM-B Model and The Behavior Change Wheel}
\begin{figure}
	\centering
	\includegraphics[width=0.5\textwidth]{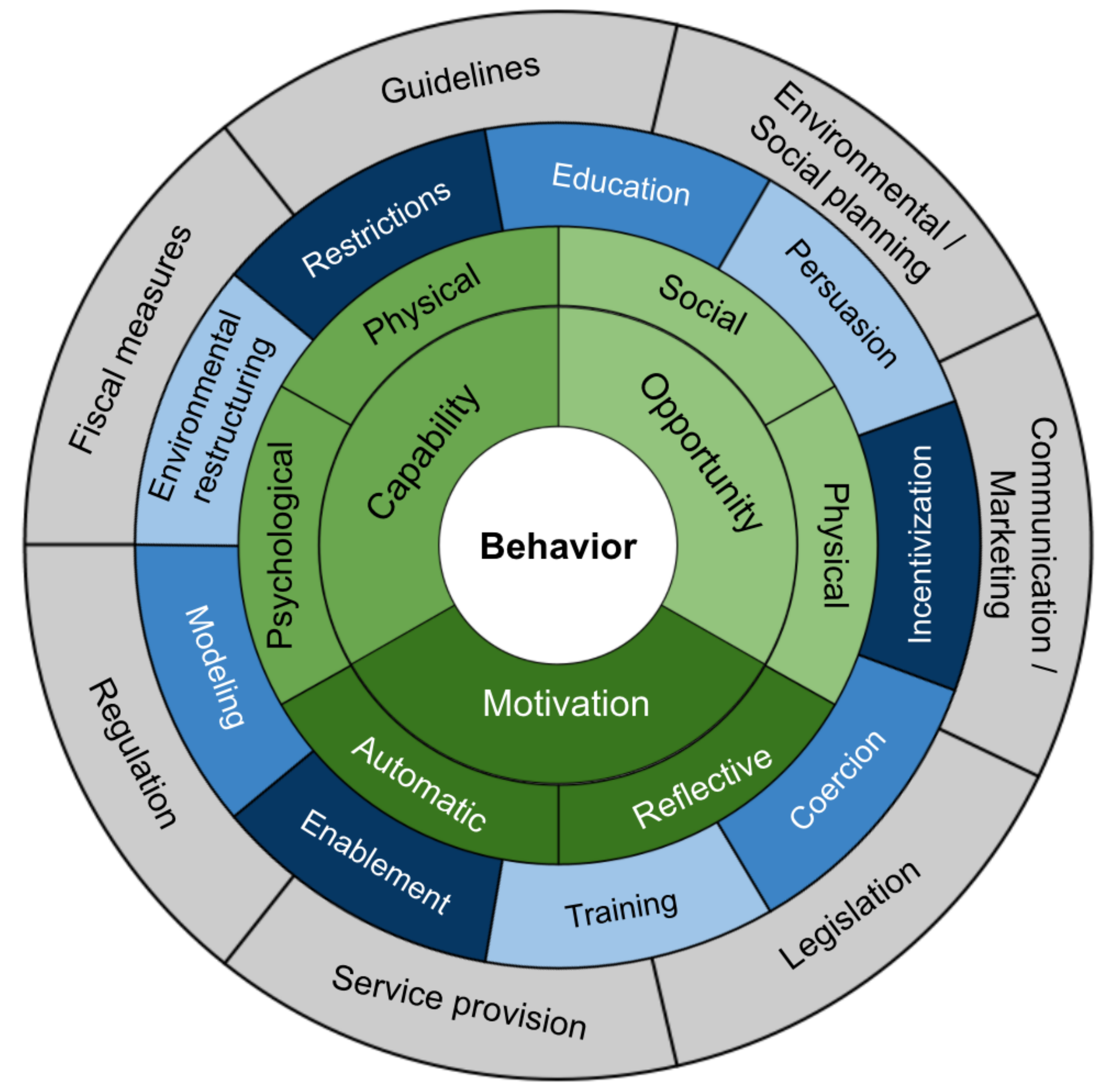}
	\caption{The Behavior Change Wheel: The  green layer refers to the COM-B model (sources of behavior), the blue layer represents the intervention functions, and the grey layer refers to the policy categories.}
	\label{fig:bcw}
\end{figure}

Michie et al. proposed \textit{the Behavior Change Wheel} (BCW) as a synthesis of 19 different theories and models of behavior change identified in a systematic literature review~\cite{michie2011}. Some of these frameworks suggest that behavior is primarily driven by beliefs and perceptions, while others significantly emphasize unconscious biases or the social environment~\cite{michie2014}. Figure~\ref{fig:bcw} shows the BCW that aims to incorporate the standard features of all these frameworks and link them to a model of behavior.  

\begin{figure}
	\centering
	\includegraphics[width=0.48\textwidth]{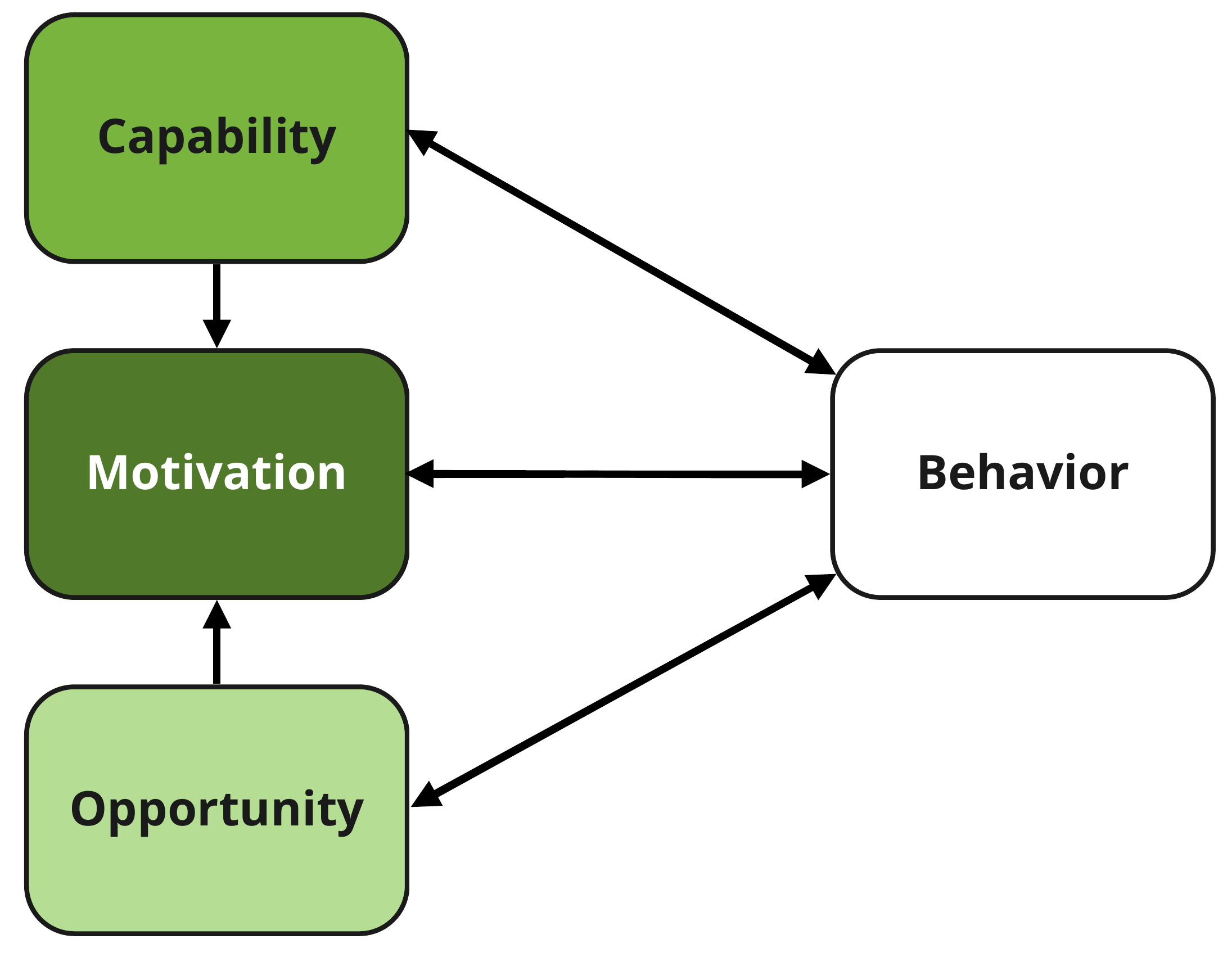}
	\caption{The COM-B model: To engage in a particular behavior, someone must have the capabilities, have the opportunity, and feel motivated to perform the target behavior more than any other competing behaviors.}
	\label{fig:comb}
\end{figure}

At the center of the BCW resides the COM-B model, shown in Figure~\ref{fig:comb}, which is composed of three vital conditions: \textbf{capability} (C), \textbf{opportunity} (O), \textbf{motivation} (M), and \textbf{behavior} (B). The COM-B model provides a clear starting point to understand behavior in a specific context, and guides the design and development of interventions. For example, according to the model, \textit{to engage in a particular behavior, someone must be \textbf{physically and psychologically capable}, have the \textbf{social and physical opportunity}, and be \textbf{motivated} to perform the target \textbf{behavior} more than any other competing behaviors}. In addition, the model presents motivation from an automatic (habits) and reflective (rational intentions) perspective. 

Figure~\ref{fig:bcw} depicts the COM-B model as the hub of the BCW; it identifies and explains the sources of the behavior, in other words, \textit{what needs to happen or change so the target behavior occurs}. Surrounding the COM-B model, two extra layers represent 9 intervention functions that will help address any issue identified in any of the COM-B model components (capability, opportunity, or motivation). Subsequently, the external layer comprises 7 policy categories that organizations can use to deliver the intervention functions. For more details, Table~\ref{tab:interventionfunction} and Table~\ref{tab:policycategory} provide the landscape of intervention functions and policy categories, including their respective definitions.

The COM-B model and the BCW have been applied in several settings, from understanding behavior change by individuals, to groups, sub-populations, and populations, and within different organizations and systems~\cite{michie2016}. For instance, Barker et al. propose a successful application of the COM-B model and the BCW to develop an intervention to promote regular, long-term use of hearing aids by adults with acquired hearing loss~\cite{barker2016}. When applying the model, the investigation exposes that behavioral planning for hearing-aid use on the side of the audiologists should be part of the routine audiological practice, which requires a complex intervention that addresses psychological, capability, physical, and social opportunity, and reflective and automatic motivation.

\subsubsection{Self-Efficacy Theory}

The Self-Efficacy Theory (SET) is an essential contribution from social cognitive theory to understand individuals' behaviors based on a self-evaluation of their abilities~\cite{bandura1978}. Bandura proposed the SET and defined it as people's beliefs about their capabilities to produce designated levels of performance that exercise influence over events that affect their lives~ \cite{bandura1994}. Therefore, high levels of self-efficacy reinforce people's convictions about their abilities to perform a task successfully~\cite{stajkovic1998}.

The SET introduces four significant sources of efficacy beliefs: \textit{mastery experiences}, \textit{vicarious experiences}, \textit{verbal persuasion}, and \textit{emotional and physiological states}~ \cite{bandura1994}. First, individual self-efficacy is boosted by having success or direct mastery experience. Additionally, observing people around us having successful experiences, especially individuals sharing similar characteristics or backgrounds, increases our beliefs that we can also achieve success by mastering the required activities. Moreover, influential people around us encourage us and raise our beliefs that we can succeed by mastering certain activities. Finally, individuals holding positive emotions are more likely to have confidence in their skills to successfully perform particular activities.

\subsubsection{Response Efficacy}
The Response-Efficacy Concept (REC) has its origins in Bandura's social cognitive theory~\cite{bandura1978}. Bandura used the term \textit{outcome expectancies} to refer to beliefs about the consequences of performing a behavior, which is the foundation of REC. Response-efficacy is defined as one's belief that acting in a specific manner is likely to mitigate threats, which is why it is generally adopted in research on fear and fear appeals~\cite{rogers1975,witte2000}. \textit{Outcome expectancies} are somewhat broader and form an essential part of people's beliefs about an attitude-object (e.g., a product, event, or behavior). The construct has its origin in Fishbein and Ajzen's expectancy-value theory~\cite{fishbein1975} and is captured under behavioral beliefs (leading to attitudes) in the theory of planned behavior~\cite{ajzen1991}. People generally develop favorable attitudes toward behaviors they believe lead to desirable consequences and form unfavorable attitudes toward behaviors they believe lead to undesirable consequences. \textit{Generalized outcome expectancies} are employed under traits such as optimism, where people generally hold that the future will turn out positively, which does not necessarily include actual behavior~\cite{carver1982}.

\subsection{Combining the Three Behavioral Models in the Context of Software Security}
To study why developers fail to adopt software security practices, we used the COM-B model as a practical diagnosis tool to highlight the capabilities, opportunities, and motivations that potentially influence developers to adopt software security practices. Additionally, we used the Self-Efficacy Theory (SET) and Response-Efficacy Concept (REC) to complement and enrich the diagnosis: SET assisted with understanding developers' beliefs regarding their capabilities and their confidence for performing software security practices (and what can influence that confidence), and REC aided in understanding how developers' perceived success in adopting security practices affects their security adoption behaviors.

The COM-B model was designed as a generic diagnosis tool. To use it in software security, we needed to adapt some of its components. The COM-B model highlights the \textit{capabilities} in terms of \textit{psychological} and \textit{physical} features. \textit{Psychological capabilities} refer to being aware of the knowledge required to perform the behavior. \textit{Physical capabilities} represent having the physical skills to conduct the behavior, for instance, having more physical strength and overcoming physical limitations. In the context of software security, since physical capabilities are not an essential component, they turned into having the \textit{technical} and \textit{non-technical} skills to adopt software security practices.

One aspect of the COM-B model emphasizes \textit{social} and \textit{physical} opportunities. On the one side, \textit{social opportunities} refer to having the opportunity afforded by interpersonal influences or social and cultural norms within the organization. On the other side, \textit{physical opportunity} implies having the resources required to perform the behavior. The term \textit{physical} turned into \textit{technical} opportunities to make it more explicit to the software security context. But, keeping the exact definition and including resources such as tools, time, and money. Finally, concerning \textit{motivations}, we used the same terminology and definitions proposed by the COM-B model. Table~\ref{tab:combcomponents} \footnote{To clarify the definitions of the COM-B model components in the context of software security, the descriptions and examples in Table~\ref{tab:combcomponents} emerged from our findings.}	 provides examples and more detail of each component definition.       

	\section{Research Method}
	\label{sec:methodology}

\begin{figure*}
	\centering
	\includegraphics[width=1.0\textwidth]{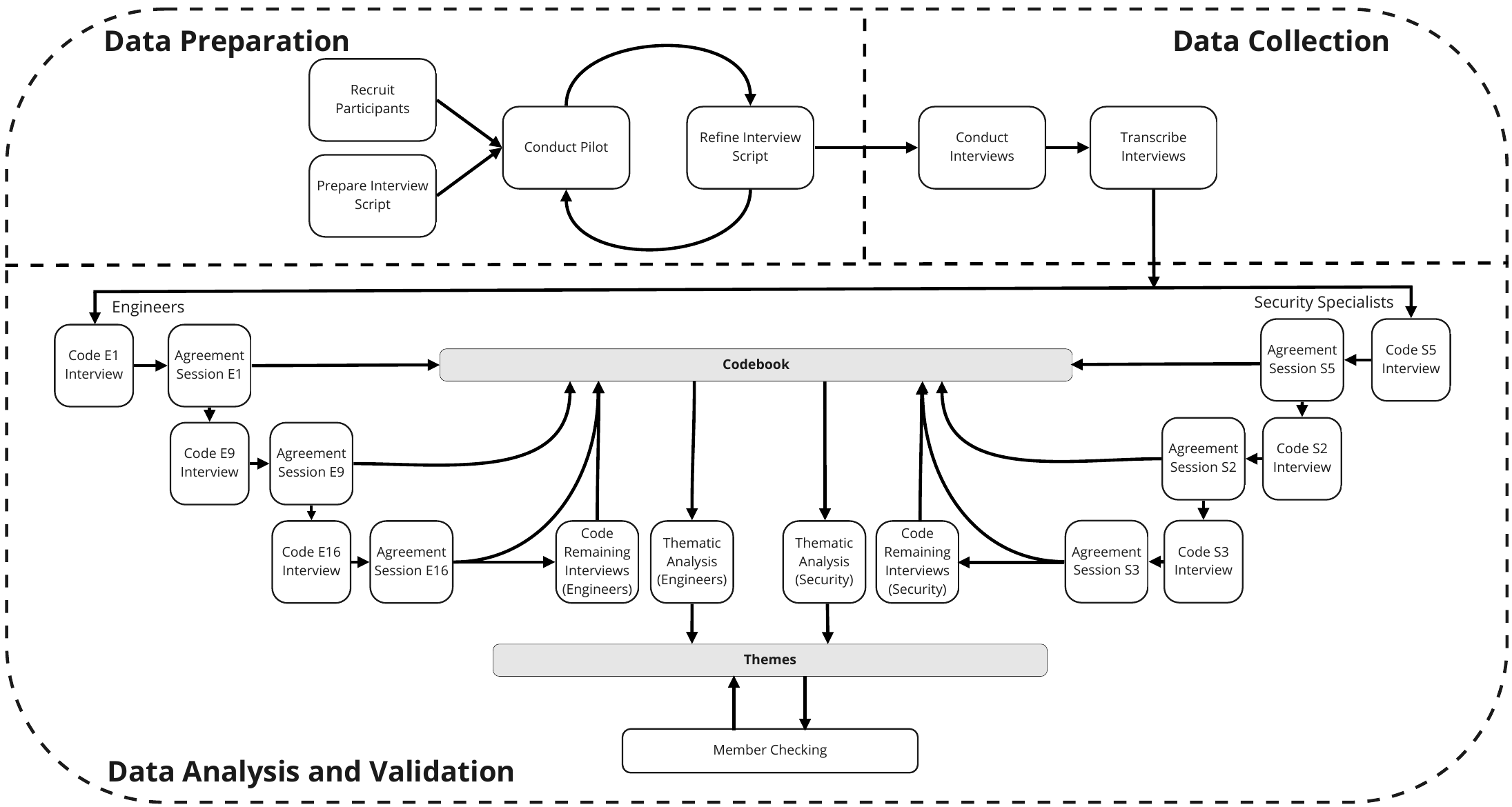}
	\caption{Research Method: Three main stages, Data Preparation, Data Collection, and Data Analysis and  Validation.}
	\label{fig:methodology}
\end{figure*}

Our study goal is to understand \textbf{\textit{what needs to happen or change, so the adoption of software security practices occurs}}. To that aim, our research methodology consists of a cross-sectional interview study~\cite{creswell2013}. 
Interview-based research is suitable for gathering thoroughly detailed participant experiences and stories~\cite{creswell2013}.
Our study design includes the following three stages: (a) Study preparation, (b) data collection, and (c) data analysis and results validation.    
Our methodology is depicted in Figure~\ref{fig:methodology}. In the following sections, we explain our methodology in detail. Private information from participants and companies has been anonymized. The authors do not have the participants' authorization to make the raw interview scripts available as they contain confidential information.

\begin{table*}[]
	\centering
	
	\caption{Profile of our participants (N=28). Companies are anonymized. MC column denotes participants who joined a member-checking session. (E1-E19) Engineering (Developer, Tech lead, Dev Ops engineer), (S1-S9) Security Specialist.}
	\resizebox{1.0\textwidth}{!}{%
	\begin{tabular}{clllrllr}
		
		\toprule
	\textbf{MC}	& \textbf{Interviewee} & \textbf{Company} & \textbf{Business} & \textbf{Company size} & \textbf{Region}  & \textbf{Role/Function}                                 & \textbf{Years of Exp.} \\
		
		\midrule     
		
&E1    & C1        & E-commerce  & 50+       &North America                     & Software Developer                  & 6             \\
&E2    & C2        & E-commerce & 1000+     &Europe           & Software Developer                            & 2             \\
*&E3    & C3        & Telecommunications & 5000+   &South America             & Software Developer                                   & 7            \\
&E4    & C4        & ERP Systems & 10000+          &North America                 & Senior Software Developer                   & 9            \\
&E5    & C5        & Beauty and Personal Care & 10+   &North America                       & Development Team Lead             & 20            \\
*&E6    & C6        & Embedded Systems & 10000+    &Europe           & Scientific Software Developer       & 2             \\
*&E7   & C7        & Education & 5000+             &Europe             & Lecturer/Software Engineer                   & 2            \\
*&E8   & C8        & Health Services & 10000+  &North America  & Senior Software Developer                            & 20             \\
*&E9   & C9        & CRM Systems & 5000+    &Europe                   & Principal Software Engineer         & 7             \\
&E10   & C10        & Booking and Rental Services & 100+   &North America      & Development Team Lead                      & 16             \\
*&E11   & C11        & Professional Design & 1000+ &Oceania & Software Developer                      & 9            \\
&E12   & C4        & ERP Systems & 10000+   &North America               & Software Developer           & 5            \\
*&E13   & C7        & Research & 5000+  &Europe   & Scientific Software Engineer                      & 2            \\
&E14   & C12        & CRM Systems & 100+  &South America   & Software Developer   & 3           \\
&E15   & C13        & Health Services & 1000+    &Europe         & Software Developer                      & 7            \\
&E16   & C14        & Video Games & 100+    &North America      & Senior DevOps Engineer         & 18             \\
&E17   & C15        & Embedded Systems & 1000+  &Europe                        & Software Engineer         & 4             \\
&E18   & C16        & Telecommunications & 50+   &North America         & VP Systems Enginnering         & 10             \\
*&E19   & C17        & Digital Publishing & 100+  &North America        & CTO         & 32             \\
\midrule
*&S1   & C18        & Health Care Services & 10000+    &North America         & Technical Product Manager         & 15             \\
&S2    & C19        & Security & 50+    &North America    & Application Security Engineer                             & 20             \\
&S3    & C20        & Financial Services & 1000+  &Europe    & Application Security Engineer                                   & 4           \\
&S4    & C21        & Security & 1000+   &Oceania     & Senior Security Developer Advocate                   & 15            \\
*&S5    & C22        & Security & 10+    &North America                      & CTO/Pen tester             & 15            \\
*&S6    & C23        & Security & 5000+  &North America             & Information Security Specialist       & 10             \\
&S7   & C24        & Telecommunications & 10000+   &North America                       & Application Security Specialist                   & 20            \\
*&S8   & C11        & Professional Design & 1000+  &Oceania   & Software and Security Engineer                            & 12             \\
&S9   & C25        & IT and Software & 10000+   &North America     & Security Program Manager         & 6             \\
		
		\bottomrule
		
	\end{tabular}%
	}
	\vspace{0.02in}
	
	\label{tab:participants}
	\vspace{-5mm}
	
\end{table*}
\subsection{Study Preparation}
Our first step was to identify potential participants for the study and define the selection criteria at this stage. Subsequently, we designed the interview script considering demographics and the different behavioral science theories we wanted to explore to understand developers' behaviors. Finally, we conducted pilots to ensure questions' comprehensibility, ensuring that sufficient time is allowed for the interviewer to conduct the interview, and reducing interviewees' cognitive load.
  
\cooltitle{Participants recruitment. }Our selection criteria was to recruit participants who had at least two years of professional work experience. Our pool of participants included two different roles, engineers and security specialists. The engineers' group consisted of software developers, tech leads,  DevOps engineers, and CTOs. The security specialists' group consisted of application security specialists, pen testers, security program managers, security developers, etc.    

The pool of interviewees from the Engineers' group came from convenience sampling~\cite{etikan2016}. The authors of this paper invited their industry contacts to participate in the study. The first author of this paper sent 42 email invitations, receiving 24 satisfactory responses to join the study. The first 5 participants were considered for piloting the interviews and refining the interview questions.  Additionally, the participants from the security specialists group came from purposive sampling~\cite{etikan2016}. The first author identified and invited software security experts using the LinkedIn platform. We sent 22 invitations, resulting in 9 of the specialists agreeing to join the study. 

In the end, our final pool of interviewees included 28 practitioners, 19 engineers (identified as E1-E19 throughout this paper), and 9 security professionals (identified as S1-S9). Our participants came from 25 different companies that work in 16 diverse industry types. The professional experience of our interviewees ranged from 2 years to 32 years, having a median of 9 years of work experience. In Table~\ref{tab:participants}, we provide more details about our participants.

\cooltitle{Interview Script Preparation. }We designed semi-structured interviews to collect our participants' experiences, stories, and challenges. Semi-structured interviews foster interviewees to freely share their experiences, enabling interviewers to explore new ideas based on the participant's answers~\cite{hove2005}. Our interview script design was guided by the COM-B model and supplemented with insights from SET and RET. A complete list of the interview questions is available in our online appendix~\cite{appendix}. The overall structure of our interview script included the following topics:

\begin{enumerate}[(a)] 
	\item the practitioners' demographics and context to understand essential aspects of the environment where software security practices' adoption (or not adoption) occurs.  
	\item how confident participants feel about their ability to perform specific software security tasks (self-efficacy).  
	\item to what extent do participants feel their adoption of software security practices impact the overall adoption in their organizations or professional network (Response efficacy Theory), and 
	\item participants' capabilities, opportunities, and motivations for adopting software security practices (COM-B model).   
\end{enumerate}     

\cooltitle{Pilot Interviews and Interview Script Refinement. }The authors conducted five pilot interviews with developers aiming to increase the comprehensibility of the interview questions and reduce participants' cognitive load during the interview. As a result, the 115 min initial interview duration was considerably reduced in each iteration, resulting in 60 minutes approximately. Two researchers were always involved during the pilots, the first author collaborating with one of the other researchers. The interviewers also asked participants their feedback regarding the questions' understandability and suggestions regarding the overall study. After each pilot interview, both researchers discussed the feedback collected and introduced the respective adjustments. After the fifth interview, the researchers agreed that the questions were mature enough. During the pilot, our interview participants pointed out that we should provide a standard definition of software security practices to avoid any potential misunderstanding at the beginning of the interviews. Additionally, our pilot participants highlighted that the interview duration should be reduced considerably, forcing us to select the most relevant questions for understanding the adoption of security practices.

\subsection{Data Collection}
The authors conducted 28 semi-structured interviews over the Zoom platform from May 31\textsuperscript{st} to July 16\textsuperscript{th} in 2021. Two researchers were involved during the interviews, one researcher leading the interview, actively asking questions and interacting with the interviewee, and the other researcher noting down relevant aspects of the participant's story and experience. We started each interview by going through our base set of questions and slightly adapting them based on the participant's role and context. For instance, we focused our questions for security professionals on their last interaction or collaboration with the engineering team instead of their personal experience as developers, which was not applicable in most cases. In addition, immediately after each interview, both researchers discussed any potential misinterpretation based on the notes taken and agreed whether the data collected reached theoretical saturation. According to Strauss and Corbin~\cite{strauss1997}, sampling should be discontinued once the data gathered no longer provides new information.     
Each interview lasted between 55 min and 75 min, and with the participant's permission, it was recorded, producing around 32 hours of recorded audio. Subsequently, recorded audios were transcribed, anonymized, and prepared for analysis.              

\subsection{Data Analysis and Findings Validation}

\begin{figure*}
	\centering
	\includegraphics[width=0.85\textwidth]{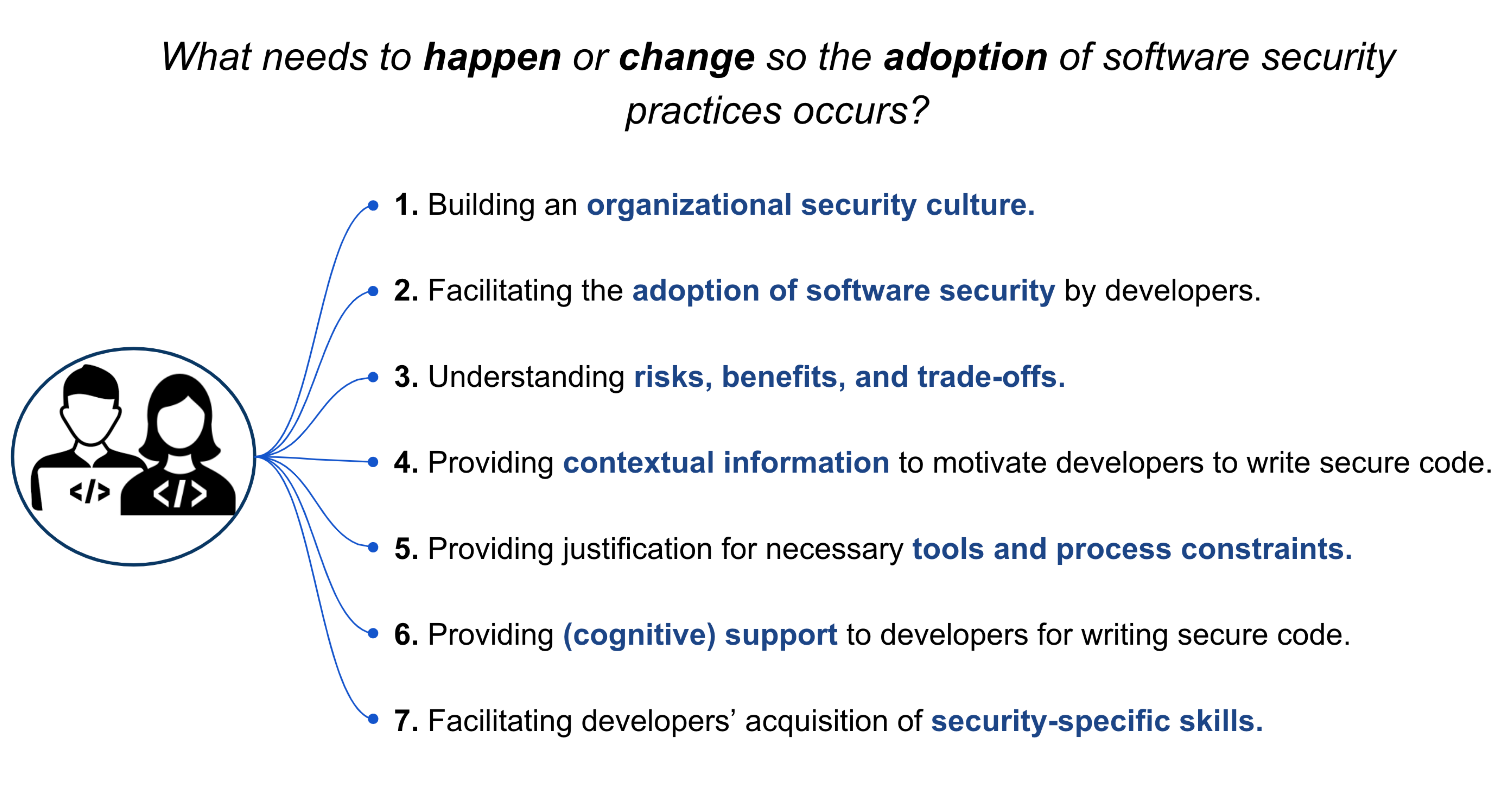}
	\caption{The \textbf{seven considerations} (categories) that \textbf{organizations} and \textbf{stakeholders} should pay attention to foster the adoption of software security practices by \textbf{developers}.}
	\label{fig:categories}
\end{figure*}  

The next stage was conducting thematic analysis to identify themes and patterns in our qualitative data~\cite{cruzes2011}. Our data analysis consisted of inductively developing codes from the transcripts and identifying themes associated with participants' adoption (or not adoption) of software security practices. We divided our data analysis into two steps. First, we analyzed data coming from the engineers' group; then, we analyzed the data from the security specialists' group. In this way, coders are not switching criteria and perspectives while analyzing the data. In addition, we followed an open coding approach~\cite{corbin1990}; during the open coding process, codes emerged and were removed or merged depending on the researchers' discussions. The discussions helped reduce any potential bias introduced by the coders.            

At the beginning of the coding phase, the first two authors coded every excerpt of the same transcript. An excerpt represents a ``dialog segment'' used as a unit of analysis associated with an interviewees' response to a question. The coding phase started with the engineers' group transcripts (E1). Both researchers coded transcript E1 independently and then calibrated their understandings of the codes in an agreement session. We defined coder agreement as an excerpt where both researchers had at least one code for the excerpt in common. Subsequently, both researchers selected another participant and continued this process iteratively until reaching an inter-rater agreement level of 85\%. Following E1, the subsequent transcripts coded were E9 and E16, respectively. Our agreement sessions involved extensive discussions on the meaning and use of the codes in our codebook, resulting in a first consensed version of our codebook. After both researchers understood each code, they started coding the remaining transcripts independently.                

To reach our study goal, we developed themes based on thematic synthesis~\cite{cruzes2011} of our coded data. We conducted 5 thematic analysis sessions with the engineers' group data, each including a different subset of participants. During the thematic analysis sessions, similarities and differences between codes were discussed and then grouped in higher-level themes. Subsequently, each theme was associated with one or more of the three components of the COM-B model (Capabilities, Opportunities, and Motivations). To validate the correct interpretation of the codes and themes creation, we counted with the feedback of an expert reviewer, the 6\textsuperscript{th} author of this paper, who helped us reduce any bias the researchers might introduce in the analysis. Additionally, to validate the association between themes and COM-B model components, we had the assistance of a domain expert in behavioral psychology, the 5\textsuperscript{th} author of this paper. 

Once the thematic analysis of the engineers' group finished, we started the open coding process for the security specialists' group. We followed a similar approach to the engineers' group. The coding phase in this group started with the first two authors coding transcript S5, then calibrating their understanding of the codes in an agreement session. Following S5, the subsequent transcripts coded were S2 and S3, respectively, reaching an average agreement of 75\%. As a result, no new code was added to the initial codebook obtained from coding the engineers' group.
Consequently, the researchers conducted two thematic analysis sessions. As a result, we added three new themes to the initial set of themes coming from the engineers' group. Similarly, in this stage, we counted with the support of an expert reviewer and a behavioral psychology expert to ensure the correct codes interpretation, themes creation, and association with the COM-B model components.  

After thematic analysis, we performed member-checking sessions~\cite{miles2014} with the study participants. We sent email invitations to all participants to join a member checking meeting to verify whether our findings are relatable to them and the context of their organizations. 12 participants accepted to join, 8 engineers and 4 security specialists. The member checking sessions lasted around 45 min and consisted of the following 5 steps: (1) the first author presenting the motivation and study participants' demographic data, (2) explicitly indicating what we want to collect in the session, (3) presenting the findings (drivers), (4) for each driver presented, discussing to what extent the driver is relatable to them and their organizations, and (5) asking for any potential driver that we overlooked in our findings. We used all participants' feedback to validate our findings and ensure the interpretation of qualitative data collected from the interviews was correct. The member-checking presentation slides and a table summarizing the feedback collected from our participants are available in our online appendix~\cite{appendix}.

\section{Findings}
\label{sec:results}

We used the COM-B model as a diagnosis tool to understand \textbf{what needs to happen or change so that developers adopt software security practices}. Based on the codes that emerged from the analysis of the interview transcripts, we associated all the excerpts from our participants' stories with the different components of the COM-B-model, \textit{capabilities}, \textit{opportunities}, and \textit{motivations}. This resulted in 33 drivers that we then grouped into seven categories based on similarity: Figure~\ref{fig:categories} shows the list of the seven categories that emerged from our study.  

In Table~\ref{tab:drivers}, we detail the entire landscape of drivers from our analysis. In the following, we describe the seven categories in detail using the components of the COM-B model associated with them, and the role of these drivers in the overall adoption of software security practices. To facilitate the description of each driver, we use the term practitioner when referring to both engineers and security specialists. As described in Section~\ref{sec:methodology}, the term \textit{engineers} includes \textit{developers, tech lead, DevOps engineers, and CTOs}, and \textit{security specialists} consists of \textit{application security engineers, pentesters, security developers}. 

\newcommand{\here}{$\bullet$}
\newcolumntype{g}{>{\columncolor{LightGray}}c}
\newcolumntype{b}{>{\columncolor{BrightGray}}c}

\begin{landscape}

\begin{table}

\caption{The \numberofdrivers drivers that emerged from our study, (E1-E19) Engineering (Developer, Tech lead, Dev Ops engineer), (S1-S9) Security Specialist. }

\begin{center}
	
\resizebox{1.37\textwidth}{!}{%
	\begin{tabular}{llbbbbbbbbbbbbbbbbbbbgggggggggrrr}
	\toprule

	&& \textbf{E1} & \textbf{E2} & \textbf{E3} & \textbf{E4} & \textbf{E5} & \textbf{E6} & \textbf{E7} & \textbf{E8} & \textbf{E9} & \textbf{E10} & \textbf{E11} & \textbf{E12} & \textbf{E13} & \textbf{E14} & \textbf{E15} & \textbf{E16} &  \textbf{E17} &  \textbf{E18} & \textbf{E19} &  \textbf{S1} & \textbf{S2} & \textbf{S3} & \textbf{S4} & \textbf{S5} & \textbf{S6} & \textbf{S7} & \textbf{S8} & \textbf{S9} & \textbf{E} & \textbf{S} & \textbf{Totals} \\
	\midrule

	\multicolumn{31}{l}{\textbf{Building an organizational security culture}} \\
\midrule

	\hspace{1mm}\textbf{D1}&Organization promoting/mandating security & \here & \here & \here & \here & \here && \here & \here & \here & \here & \here & \here & \here & \here & \here & \here & \here & \here & \here & \here & \here & \here & \here & \here & \here & \here & \here & \here & 18 & 9 & 27 \\
	\hspace{1mm}\textbf{D2}&Prioritizing security practices && \here & \here && \here && \here & \here & \here && \here & \here & \here & \here & \here & \here & \here & \here && \here & \here & \here & \here & \here & \here & \here & \here & \here & 14 & 9 & 23 \\
	\hspace{1mm}\textbf{D3}&Having a security-specific role filled & \here & \here &&&&&& \here & \here & \here & \here & \here && \here & \here & \here && \here && \here & \here & \here && \here & \here & \here & \here & & 11 & 7 & 18 \\
	\hspace{1mm}\textbf{D4}&Overcoming the resistance to change &&&& \here &&&&&&\here &&&&&&&& \here &\here && \here &&&& \here &&&  & 4 & 2 & 6 \\
	\hspace{1mm}\textbf{D5}&\thead[l]{Fostering collaboration between engineering and security teams} & \here & \here && \here &&& \here & \here & \here & \here & \here & \here && \here & \here & \here & \here &&& \here && \here & \here && \here & \here & \here & \here  & 13 & 7 & 20 \\
	\hspace{1mm}\textbf{D6}&\thead[l]{Awareness of the social perception of security adoption in one's own\\organization and professional network} & \here & \here & \here & \here & \here & \here & \here & \here & \here & \here & \here & \here & \here & \here & \here & \here & \here & \here & \here & \here & \here & \here & \here & \here & \here & \here & \here & \here & 19 & 9 & 28 \\
	\hspace{1mm}\textbf{D7}&\thead[l]{Providing awareness of external incentives and compliance} && \here &&& \here & \here &&&&&&&&& \here &&&&&&&& \here & \here & \here & \here & \here & & 4 & 5 & 9 \\

	\midrule
	\multicolumn{31}{l}{\textbf{Facilitating the adoption of software security by developers}} \\
	\midrule

	\hspace{1mm}\textbf{D8}&Shaping developer's attitude towards security & \here & \here & \here & \here & \here &&&&& \here & \here & \here &&&& \here & \here && \here && \here & \here && \here & \here & \here & \here & \here & 11 & 7 & 18 \\
	\hspace{1mm}\textbf{D9}&Tool awareness & \here & \here & \here && \here && \here && \here & \here &&& \here &&&&&& \here & \here & \here & \here & \here & \here &&& \here & \here  & 9 & 7 & 16 \\
	\hspace{1mm}\textbf{D10}&Standard guidelines geared at developers &&& \here &&& \here && \here & \here & \here && \here &&& \here & \here &&&& \here & \here & \here & \here & \here &&&& & 8 & 5 & 13 \\
	\hspace{1mm}\textbf{D11}&Reduction of system complexity & \here & \here &&& \here & \here &&&&& \here &&& \here & \here &&& \here &&& \here &&&&&&& & 8 & 1 & 9 \\
			
	\midrule
	\multicolumn{31}{l}{\textbf{Understanding risks, benefits, and trade-offs}} \\
	\midrule
	
	\hspace{1mm}\textbf{D12}&Awareness of potential risks and security incidents & \here & \here & \here & \here & \here & \here & \here & \here & \here & \here & \here & \here & \here & \here & \here & \here & \here & \here & \here & \here && \here & \here && \here & \here & \here & \here  & 19 & 7 & 26 \\
	\hspace{1mm}\textbf{D13}&Learning from actual incidents &&&&&&& \here &&& \here & \here && \here & \here &&& \here &&& \here &&& \here && \here &&& & 6 & 3 & 9 \\
	\hspace{1mm}\textbf{D14}&Fear of non-adoption consequences &&&&& \here & \here & \here &&& \here & \here & \here & \here & \here & \here && \here & \here & \here & \here & \here && \here & \here && \here & \here & \here & 12 & 7 & 19 \\
	\hspace{1mm}\textbf{D15}&Knowledge of benefits &&& \here & \here & \here & \here & \here & \here && \here & \here & \here & \here & \here & \here && \here & \here & \here & \here & \here & \here & \here & \here & \here & \here & \here & \here & 15 & 9 & 24 \\

	\midrule
	\multicolumn{31}{l}{\textbf{Providing contextual information to motivate developers to write secure code}} \\
	\midrule
	
	\hspace{1mm}\textbf{D16}&Promoting a customer satisfaction/protection mindset &&&&&&&&& \here &&& \here &&& \here && \here &&& \here &&& \here & \here &&&& & 4 & 3 & 7 \\
	\hspace{1mm}\textbf{D17}&\thead[l]{Awareness of the influential role of the industry type in developers'\\ disposition towards security compliance} & \here &&& \here &&&&& \here & \here && \here & \here & \here & \here & \here &&& \here &&&& \here & \here && \here & \here & & 10 & 4 & 14 \\
	\hspace{1mm}\textbf{D18}&\thead[l]{Awareness of developers' perceptions of the need for software security\\based on application characteristics}&&& \here & \here && \here & \here & \here && \here & \here & \here && \here & \here & \here & \here & \here &&&& \here & \here && \here & \here & \here & \here  & 13 & 6 & 19 \\
	\hspace{1mm}\textbf{D19}&\thead[l]{Aligning the perspective of what "good enough" security means} && \here & \here && \here & \here & \here & \here & \here & \here & \here & \here & \here && \here & \here & \here & \here & \here &&& \here & \here & \here & \here & \here & \here & \here & 16 & 7 & 23 \\

	\midrule
	\multicolumn{31}{l}{\textbf{Providing justification for necessary tools and process constraints}} \\
	\midrule
	
	\hspace{1mm}\textbf{D20}&Consideration of tool constraints on developers' autonomy & \here & \here && \here &&&&& \here &&& \here & \here & \here & \here && \here &&&&&& \here &&&&& \here & 9 & 2 & 11 \\
	\hspace{1mm}\textbf{D21}&Awareness of developers' perception of security-imposed restrictions & \here && \here & \here & \here & \here &&& \here && \here & \here & \here & \here & \here & \here && \here &&&&& \here & \here & \here & \here & \here & \here & 13 & 6 & 19 \\

	\midrule
	\multicolumn{31}{l}{\textbf{Providing (cognitive) support to developers for writing secure code}} \\

	\midrule
	
	\hspace{1mm}\textbf{D22}&Availability of reminders, i.e., checklists, dashboards, etc. && \here & \here & \here & \here & \here & \here &&& \here & \here && \here & \here && \here && \here & \here & \here &&&& \here & \here && \here & \here  & 13 & 5 & 18 \\
	\hspace{1mm}\textbf{D23}&Improving the usability (complexity reduction) and accuracy of security tools & \here & \here &&&&&&& \here && \here & \here &\here &&&&&&&&&&&&&&& & 6 & 0 & 6 \\
	\hspace{1mm}\textbf{D24}&Reducing the effort required to learn or apply security & \here & \here & \here & \here & \here & \here & \here && \here & \here & \here & \here & \here & \here & \here & \here & \here & \here & \here & \here & \here & \here & \here & \here & \here & \here & \here & \here & 18 & 9 & 27 \\
	\hspace{1mm}\textbf{D25}&Integrating tools into the development workflow &&&&&&&&&&& \here && \here & \here & \here &&& \here && \here && \here & \here &&& \here & \here & \here & 5 & 6 & 11 \\			

	\midrule
	\multicolumn{31}{l}{\textbf{Facilitating developers' acquisition of security-specific skills}} \\
    \midrule
    
	\hspace{1mm}\textbf{D26}&Accessibility to learning resources & \here & \here & \here & \here & \here & \here &&& \here & \here & \here & \here && \here && \here && \here &&& \here & \here & \here & \here & \here &&& \here  & 13 & 6 & 19 \\
	\hspace{1mm}\textbf{D27}&Using security practices as learning tools & \here && \here &&&&&& \here & \here && \here && \here &&&&&&& \here & \here & \here &&&&& & 6 & 3 & 9 \\
	\hspace{1mm}\textbf{D28}&Providing security education & \here & \here && \here &&& \here & \here & \here & \here & \here & \here & \here & \here & \here && \here & \here & \here & \here & \here & \here && \here & \here & \here & \here & \here & 15 & 8 & 23 \\
	\hspace{1mm}\textbf{D29}&Fostering hands-on learning/self-learning/osmosis & \here & \here & \here && \here && \here &&& \here & \here & \here && \here &&& \here &&&& \here & \here && \here & \here & \here & \here & \here & 10 & 7 & 17 \\
	\hspace{1mm}\textbf{D30}&Creating and participating in communities of practice & \here && \here & \here & \here &&& \here & \here & \here && \here && \here &&&& \here & \here & \here & \here & \here & \here & \here & \here & \here & \here & \here & 11 & 9 & 20 \\
	\hspace{1mm}\textbf{D31}&Having non-technical skills & \here & \here & \here & \here & \here & \here & \here & \here & \here & \here & \here & \here & \here & \here & \here & \here & \here & \here & \here & \here & \here & \here & \here & \here & \here & \here & \here & \here & 19 & 9 & 28 \\
	\hspace{1mm}\textbf{D32}&Confidence in their technical abilities & \here & \here & \here & \here & \here & \here & \here & \here & \here & \here & \here & \here & \here & \here & \here & \here & \here & \here & \here & \here && \here &&& \here & \here & \here & \here & 19 & 6 & 25 \\
	\hspace{1mm}\textbf{D33}&Awareness of necessary security skills & \here &&&&& \here & \here && \here && \here & \here & \here & \here & \here && \here & \here && \here & \here & \here & \here & \here & \here & \here & \here & \here & 11 & 9 & 20 \\

	\bottomrule
	\end{tabular}%
   }

\vspace{0.02in}

\end{center}

\label{tab:drivers}

\end{table}

\end{landscape}

\subsection{Building an organizational security culture}
\begin{figure}
	\centering
	\includegraphics[width=0.5\textwidth]{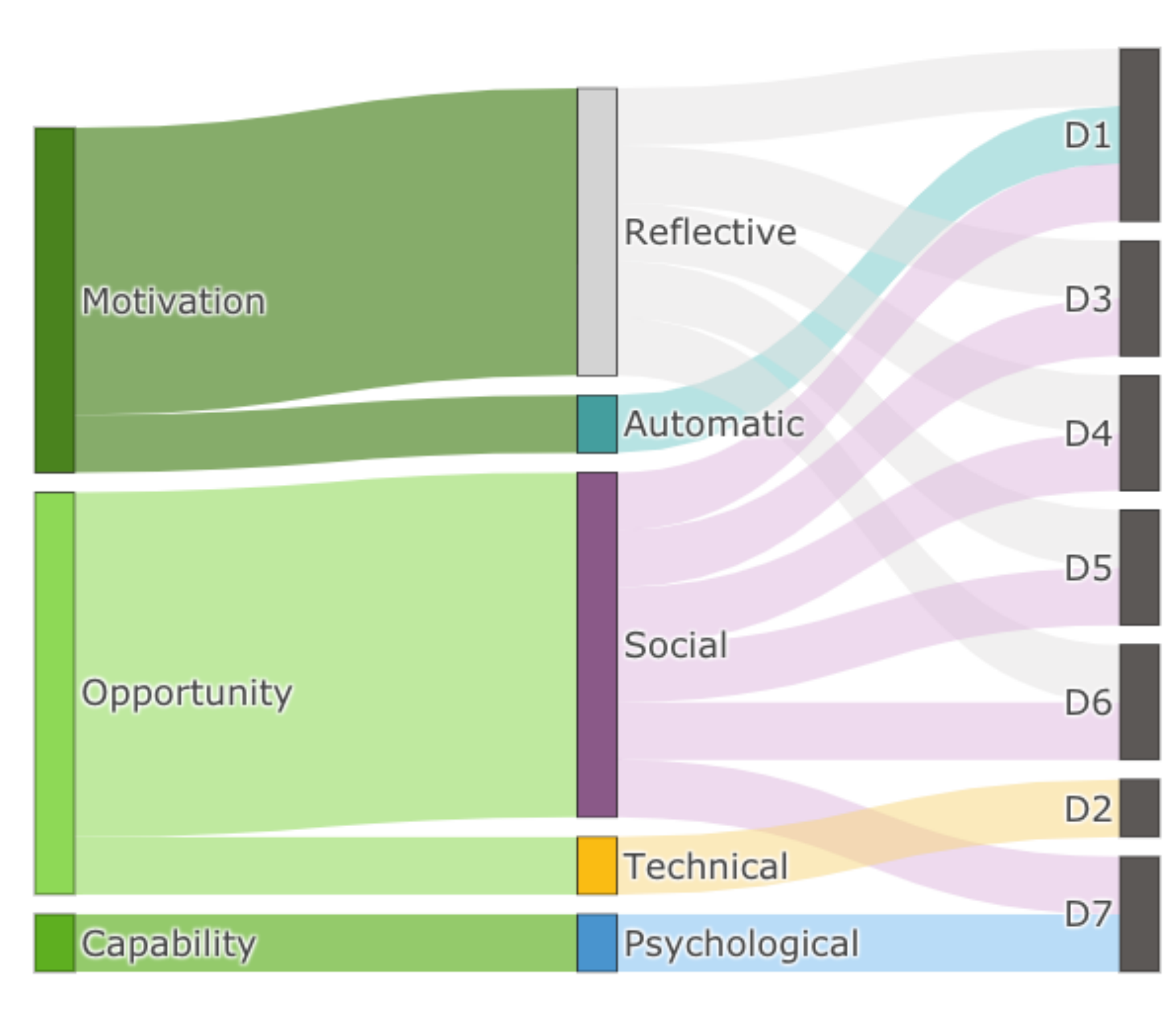}
	\caption{\textbf{Building an organizational security culture:} 
		\textbf{D1:} Organization promoting/mandating security,
		\textbf{D2:} Prioritizing security practices, 
		\textbf{D3:} Having a security-specific role filled, 
		\textbf{D4:} Overcoming the resistance to change, 
		\textbf{D5:} Fostering collaboration between engineering and security teams, 
		\textbf{D6:} Awareness of the social perception of security adoption in one's own organization and professional network, and
		\textbf{D7:} Providing awareness of external incentives and compliance.
	}
	\label{fig:category1}
\end{figure}

Practitioners consider culture an essential driver to boost the adoption of security practices in an organization. Organizations that promote or mandate security usually exhibit a good security posture by having a role such as a security champion or a security team in the company. Figure~\ref{fig:category1} shows the relationships between the COM-B model components and drivers.  

\cooltitle{D1: Organization promoting/mandating security.}
Organizations should carefully pay attention to their influential role in developers' adoption of software security practices. We used the COM-B model as a diagnosis tool to highlight the \textit{social opportunities}, \textit{reflective motivations}, and  \textit{automatic motivations} of the behaviors related to the adoption of software security practices. \textit{Social opportunities} represent the organization's norms and values influencing developers' attitudes towards security. \textit{Reflective motivations} indicate how those norms and values encourage developers to reflect on their practices and adopt new habits, and \textit{automatic motivations} indicate how developers' emotions react to those norms and values. 

\cooltitlecomb{Social opportunities: }Practitioners indicated that organizational culture highly influences developers' attitudes towards security. Developers are willing to adopt security practices if organizations give them the proper time and opportunity to learn and apply security in their workflow. For instance, in the words of E9, ``\textit{Adopting security requires support from the organization by facilitating the solution of a security issue.}'' 
Sometimes organizations might treat security as a second-class citizen. For example, E16 emphasizes: ``\textit{Making it work is more important in my organization than doing it securely.}''   

\cooltitlecomb{Reflective motivations: }Engineers revealed that organizations interested in embracing security as part of their culture care about developers' motivations for adopting software security practices. 
Strategies to disseminate security guidelines and practices across the organization influence developers' motivations for adopting software security practices. These strategies are perceived by developers as a good sign of proactiveness towards security, and they make practitioners reflect 
on their practices and align them with the rest of the organization. For example, E16 said: ``\textit{The company is pushing for security. Sending in the monthly newsletter what new security rules have been put in place.}''

\cooltitlecomb{Automatic motivations: } Engineers experience frustration when organizations, particularly management and other departments, do not treat security as a first-class citizen, making it difficult for developers to prioritize security over other tasks. Sometimes management pushes developers to ensure a secure product despite time pressure issues and several barriers their organizations introduce that can delay the delivery of a feature or product. These delays might be caused by waiting for feedback or approval from a security team or legal and privacy department. For instance, E7 pointed out: ``\textit{So currently, we are annoyed by the slowness of the organization, so there is a legal department and privacy department, and they need to evaluate our system in terms of security, and they're just really slow in doing so.}''

\cooltitle{D2: Prioritizing security practices.} 
Not all organizations prioritize security in the same way. Depending on how critical the data managed by the application is and the resources available, organizations, if needed, will introduce different security practices across the development pipeline to ensure a secure product. Shifting security to the left, at the design stage, or leaving security to later stages during testing or operational stages are standard practices seen in the industry.

\cooltitlecomb{Technical opportunities: } Organizations that prioritize security provide developers with opportunities to adopt security practices by enabling them with the appropriate resources and processes to ensure a flawless product. Typically, organizations introduce several security inspection mechanisms during the software development pipeline to rigorously identify potential vulnerabilities in the supply chain and ensure that malicious hackers cannot compromise personally identifiable information from customers. For instance, E18 highlighted: ``\textit{Security is essential for us ... There's a considerable amount of privacy work that you have to do upfront in terms of getting the consent, ensuring that you're handling the data properly ... We do regular audits and reviews from the privacy side that covers the security of the application.}''

\cooltitle{D3: Having a security-specific role filled.}
Practitioners agreed that the presence of a security role within the  organization can promote, maintain, and enforce security practices. Concerning the COM-B model, we identified that organizations employing a security role provides them with  \textit{social opportunities} and \textit{reflective motivations}. \textit{Social opportunities} are relevant due to the influential role of a security specialist in adopting software security practices within the organization. Additionally, \textit{reflective motivations} represent how the presence of a security-specific role pushes developers to think about the security implications of their technical decisions.

\cooltitlecomb{Social opportunities: } Practitioners highlighted the importance of having a security role in the organization. An organization that promotes a software security culture designates a specific position for security matters. In addition, developers are willing to adopt software security practices if organizations allow them to shadow security specialists to learn and understand how security issues are handled and patched. For instance, E1 stated: ``\textit{I will learn from them, see how people fix the problem, how they prioritize the problem, and  that's one thing the company is really good at, letting you see what happens behind the scenes so you can get a good understanding of it. And eventually, that will translate into me working on more and more security-related tickets.}''

\cooltitlecomb{Reflective motivations: }Security specialists can facilitate organizational discussions about security practices, the consequences of security issues, and potential threats to the system. Hence, developers will be aware of more profound implications of security issues and, as a result, will understand the rationale behind security guidelines and feel motivated to incorporate them into their software development workflow. For example, E10 pointed out:  ``\textit{Having a security role in the organization, probably fosters a better environment for adopting security practices. I think that if you have that kind of security experts scattered about your organization, I think that brings everyone's security knowledge level up.}''

\cooltitle{D4: Overcoming the resistance to change.}
Practitioners also implied that in some particular cases, developers might be reluctant to adopt new engineering practices, specifically security, because software security requires extra effort in their regular workflow---a scenario that developers are not willing to accept and a highly compelling reason behind their lack of interest.     

\cooltitlecomb{Social opportunities: }Engineers perceived that adopting security practices require convincing senior management that security is necessary and involves a cultural change where security is promoted from top management to the rest of the organization. In addition, it requires changing a collective mindset from just delivering everything as fast as possible to building a trustworthy and secure product. For example, E19 emphasized: ``\textit{For adopting security, executive sponsorship is going to be the most important, making sure that you've got the backing. Then it's just a standard change management practice. So make sure that you've got the executive sponsorship that understands the value of security and will fight for it. That's probably the most important.}''   

\cooltitlecomb{Reflective motivations: }In particular cases, engineers are reluctant to change their usual engineering practices and adopt new ones. Some developers with extensive experience in development might have a negative attitude towards adopting security practices. For instance, E18 pointed out: ``\textit{Some people just might like their workflow so much, and they just don't feel like much of adopting security. They feel that the type of stuff they're working on, or the type of code they are developing, wouldn't be better by adding security.}''

\cooltitle{D5: Fostering collaboration between engineering and security teams.}
Practitioners highlighted the need to provide effective mechanisms to enhance collaboration between engineering and security teams. Typically, when software security is mandated on organizations, developers negatively perceive a security team's involvement in technical decisions across the development pipeline. For instance, some negative effects as perceived by developers are: extra work, delays, rework, or conflicting perspectives to prioritize and solve an issue.     

\cooltitlecomb{Social opportunities: }According to security specialists, most developers perceive security teams as \textit{the carrier of bad news}. The ones who will notify them about their security mistakes. Especially when a security vulnerability is identified or as soon as an application is attacked, for instance, S4 emphasized: ``\textit{Security team is usually going to be traditionally the ones that engage with devs to try and find out how to fix something that they've identified in a particular deployment.}''

\cooltitlecomb{Reflective motivations: }Security specialists perceived it is essential to work in close collaboration with developers. Developers will be encouraged to follow security practices if they feel supported by security experts during security-related tasks, facilitating the learning process and reducing the effort to apply security in their regular workflow. For example, S3 highlighted: ``\textit{For collaborations to happen is vital to understand one another, like understanding how development works and how security works. Because when one doesn't know how the other works, they will assume things and do things on how they best understand. So the other one can be put aside by mistake, and the collaboration can fail.}''

\cooltitle{D6: Awareness of the social perception of security adoption in one's own organization and professional network.}
Most engineers perceived that adopting software security practices is a collaborative effort that's strongly influenced by their professional community. For instance, developers being aware of their peers' efforts to adopt software security practices provides an opportunity to join a collective effort.  

\cooltitlecomb{Social opportunities: }Practitioners indicated that having people around them adopting security practices helps them follow the security procedures and reminds them why to invest extra effort for adopting those practices. For instance, E15 pointed out: ``\textit{What helps you as an external motivation to the team is having people around you that have the procedures in place. And remind you why we have those procedures, because some people don't like that, but I think it's important, so I don't mind putting extra effort into it.}''

\cooltitlecomb{Reflective motivations: }The overall adoption within the organization significantly influences developers' perception of software security practice adoption. Not having people within the organization adopting security will considerably diminish developers' motivations to adopt security. For example, E16 pointed out: ``\textit{if nobody else takes it seriously, I'll never take it seriously. If it's not part of the culture, if it's just one guy saying security, security, security, then people will do the bare minimum to adopt security, which might be better than nothing. Still, it needs to be a part of everyone.  Everybody has to care about it for you to feel like you're making secure software.}''

\cooltitle{D7: Providing awareness of external incentives and compliance.}
Organizations aware of external incentives, such as potential government subsidies for security testing, particularly among start-ups, represent an excellent opportunity to adopt software security practices. Additionally, developers perceive that organizations should promote awareness among developers of external regulations and compliance from the beginning of the software project, explicitly highlighting the practices, technical considerations, and justification or rationale behind the compliance. 

\vspace{10px}

\cooltitlecomb{Psychological capabilities: }Practitioners perceived it is vital for organizations to bring awareness of the compliance standards that regulate data protection and privacy, such as GDPR~\footnote{https://gdpr-info.eu/}, PCI~\footnote{https://www.pcisecuritystandards.org/}, and HIPPA~\footnote{https://www.hhs.gov/hipaa/index.html}. To achieve compliance, organizations reinforce specific procedures to ensure the privacy and security of all customers' data managed by the application. For  instance, some standard security practices observed are encryption mechanisms and regular penetration testing activities. Additionally, external incentives play a crucial role in adopting software security practices. For example, government incentives such as IRAP~\footnote{https://nrc.canada.ca/en/support-technology-innovation/about-nrc-industrial-research-assistance-program} allow startup organizations in Canada to subsidize 24 hours of security testing. For example, S5 highlighted: ``\textit{We work with many startups that the government subsidizes their testing. So I think that's an excellent option for startups. It's only a three-day engagement, but it's a perfect start to get an idea of where they're at. If we find that they have many input validation issues, that shows that there's something wrong with the process that needs to be addressed. If we see many configuration issues, you know, you might get insight into where they're having problems. So that can give them kind of focused advice.}''    
    
\cooltitlecomb{Social opportunities: }Practitioners perceived organizations dealing with sensitive customer data or safety-critical systems adopt software security practices for compliance reasons. However, without the support of the business, the adoption will not occur. For example, S7 pointed out: ``\textit{Security is a thing that they need to do for compliance or contractual reasons. But if the business does not support security concerns, they will not adopt it. Some organizations might be willing to sacrifice security and take penalties on contracts if that works out business-wise in their favor.}''

\subsection{Facilitating the adoption of software security by developers}
\begin{figure}
	\centering
	\includegraphics[width=0.5\textwidth]{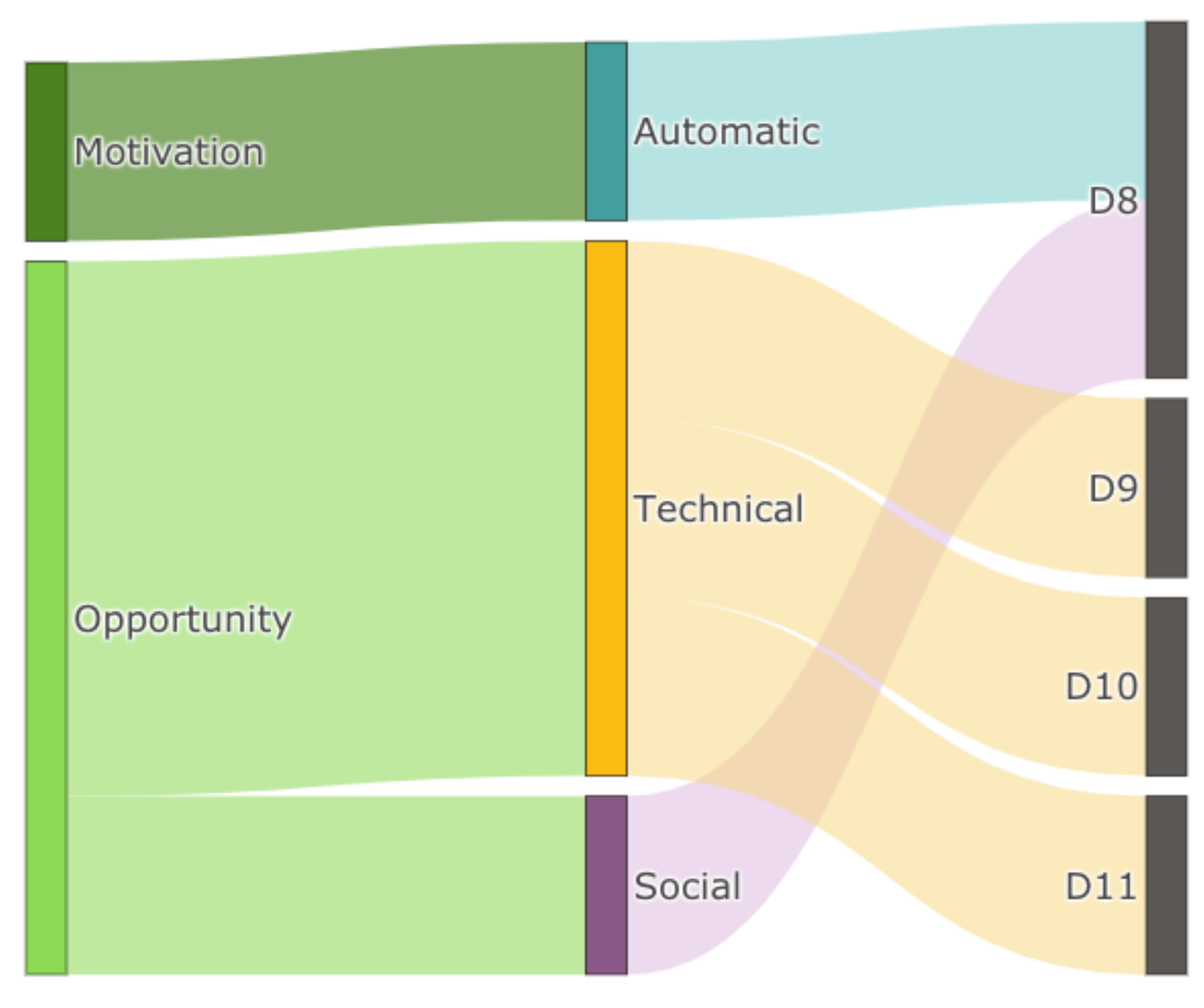}
	\caption{\textbf{Facilitating the adoption of software security by developers:} 
	\textbf{D8:} Shaping developers' attitudes towards security, 
	\textbf{D9:} Tool awareness, 
	\textbf{D10:} Standard guidelines geared at developers, and
	\textbf{D11:} Reduction of system complexity. 
	}
	\label{fig:category2}
\end{figure}

Developers pointed out that they usually do not see the immediate benefits of adopting security practices but instead the adverse effects of introducing security practices that affect their development pipeline, e.g., delays. Based on our analysis, we describe the organization's crucial role in easing the adoption of software security by developers. Figure~\ref{fig:category2} shows the relationships between the COM-B model components and the drivers from our analysis.

\cooltitle{D8: Shaping developers' attitudes towards security.}
Software practitioners indicated that it is easier for developers to identify the disadvantages of adopting security instead of the immediate benefits, introducing negative attitudes towards its adoption. Developers perceived that organizations have a relevant role in shaping developers' attitudes towards security by introducing security as a ``fun activity'' and providing opportunities, e.g., through gamification, to adopt and learn software security practices effortlessly. 

\cooltitlecomb{Social opportunities: }Organizations play a crucial role in fostering opportunities to shape developers' attitudes towards security. Organizations that provide the conditions to get developers involved in security incident fixing processes positively affect developers' attitudes. Practitioners perceived this scenario as a promising chance to learn new technical skills as part of their regular development work. For example, E1 said: ``\textit{That's one of the things the company is good at, letting you see what happens behind the scenes so you can get a good understanding of it. And eventually, that will translate into me working on more and more security-related tickets.}'' 

\cooltitlecomb{Automatic motivations: }Developers working in organizations that promote software security practices often perceive security as challenging but rewarding work. In addition, developers with a positive attitude towards security feel highly motivated to adopt security practices to protect customers, users, and the company. For instance, E1 highlighted: ``\textit{I think of security as a chess game. I play one side; the attackers play the other side. It can be quite challenging, it's hard work, but it's rewarding in the end, so my motivation is to protect people and protect the company.}'' Additionally, developers appreciate organizations' effort to introduce security as a fun activity. For example, E2 pointed out: ``\textit{My team did the hack your own product day, and that's a fun experience.}''

\cooltitle{D9: Tool awareness.}
Organizations that aim to facilitate the adoption of software security practices typically provide the technical resources developers require to adopt it, specifically specialized security tools. Developers find it easier to adopt security if they are able to access appropriate tools.     

\cooltitlecomb{Technical opportunities: }Engineers believe that organizations eager to create a culture around security should actively introduce tools as part of the teams' discussions. Developers found valuable tools that can be easily integrated into their development workflow and notify them of issues or alert them of potential risks. For instance, E5 highlighted: ``\textit{Tools involved in the process, that make a lot of sense, providing educational resources where needed. But of course, the challenge with some tools is that you don't realize them till later. The earlier we can identify that stuff, the better.}''

\cooltitle{D10: Standard guidelines geared at developers.}
Engineers often perceive that guidelines focused on software security are quite abstract and overly complicated for their particular information needs. Organizations play a significant role in customizing security guidelines to a developer's context and workflow. Developers will then be eager to use those security guidelines, perceive their benefit, and apply them to their regular practice.    

\cooltitlecomb{Technical opportunities: } Practitioners perceived that most security guidelines are not developer friendly. There is still much work needed to make those guidelines comprehensible by all security stakeholders, particularly developers. For example, S3 emphasized: ``\textit{We have a considerable fragmentation in the current application security culture. So let's say you're a developer and I'm a tester. I can't communicate with you through a unified standard. I can use CWE, CVSS, anything you can think of. And the developer will not understand what I'm talking to them.}''

\cooltitle{D11: Reduction of system complexity.}
Practitioners recognized that applications evolve innately, increasing in size and complexity, making maintenance and security management harder. Developers acknowledge  organizations' effort to simplify the adoption of security by abstracting security to a specific layer/component/service in the application, e.g., applying separation of concerns.

\cooltitlecomb{Technical opportunities: }Inevitably, systems grow continuously, making their management more complex from a security point of view. Developers perceived useful when organizations simplified security from development by fostering the adoption of security frameworks, and protocols or treating security as a core feature in the application under the management of specialized teams. For example, E11 highlighted: ``\textit{Many security details are abstracted away from development. So they are just part of the policies or the plan. And occasionally, you can contact the security team to suspend specific user permissions, but from my point of view, it's just very well embedded in the process. We don't need to know about the details, but we need to know who needs to be contacted and which requirements need to be fulfilled from a security standpoint.}''

\subsection{Understanding risks, benefits, and trade-offs}
\begin{figure}
	\centering
	\includegraphics[width=0.5\textwidth]{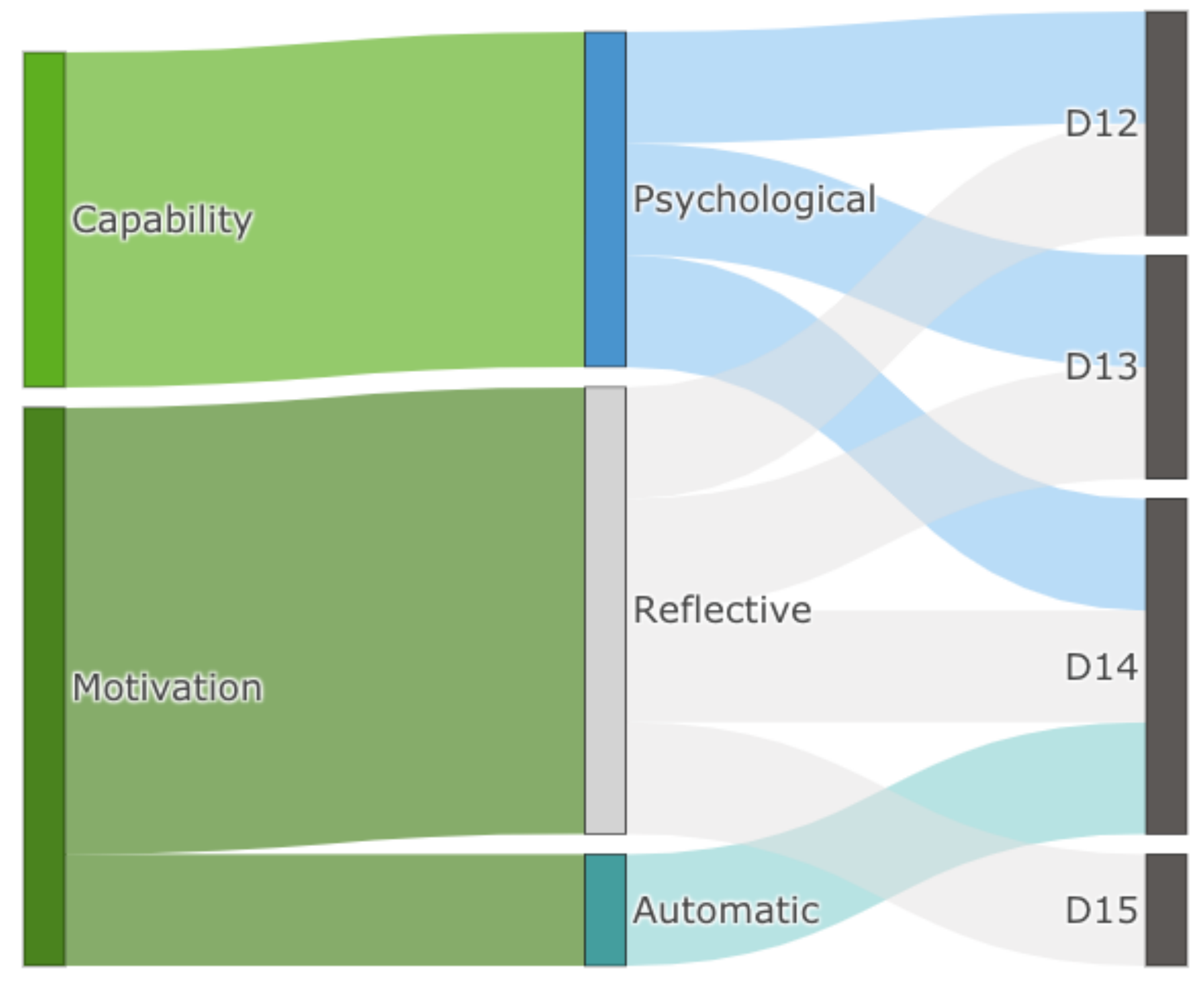}
	\caption{\textbf{Understanding risks, benefits, and trade-offs:} 
	\textbf{D12:} Awareness of potential risks and security incidents, 
	\textbf{D13:} Learning from actual incidents, 
	\textbf{D14:} Fear of non-adoption consequences, and
	\textbf{D15:} Knowledge of benefits.
	}
	\label{fig:category3}
\end{figure}

Practitioners perceived that bringing awareness about security risks, providing information about the consequences of not adopting security practices, and discussing the benefits of building security into the development pipeline will incentivize them to adopt security. Furthermore, being aware of the trade-offs of adopting software security practices will help developers maintain their positive attitude towards security and continuous interest in inspecting security defects in their products. Figure~\ref{fig:category3} shows the relationships between the COM-B model components and drivers.  

\cooltitle{D12: Awareness of potential risks and security incidents.}
Being aware of existing threats and vulnerabilities that software products face is one of the critical motivations for developers to adopt security practices. In addition, by being aware of what risks other organizations face, developers can identify what is required to secure an application and prioritize activities accordingly.

\cooltitlecomb{Psychological capabilities: }Developers often use security news on the Internet to learn about security incidents and be aware of how similar organizations have been exploited as a measure to prevent those situations from happening in their organizations. For instance, E1 stated: ``\textit{Just digging around, seeing what information I could learn about on the Internet and see what was going on. You read many security news articles about people getting hacked, and I was just curious to see how they did it.}''

\cooltitlecomb{Reflective motivations: }Developers perceived that when security is a significant concern in the business they belong to, they constantly gather information about security exploits in the industry. Therefore, this situation pushes developers to carefully examine their security practices to ensure a secure product for their customers. For example, E19 emphasized:  ``\textit{It's hard to get a computer science degree or a software engineering degree and not be mindful and aware of security. People in this field tend to keep up with the latest news related to technology, both out of personal interest and professional interest. It's pretty much impossible to ignore what's happening with all the security breaches occurring all the time. So it's part of the consciousness of most developers.}'' Additionally, E7 pointed out: ``\textit{We do care a lot about security, especially we started caring a bit more about security after some similar organizations got hacked. So we decided to look closely and see if we are doing everything up to standards.}''

\cooltitle{D13: Learning from actual incidents.}
Practitioners indicated that having the experience of being hacked is one of the best learning resources for adopting software security practices. For instance, studying how security exploits happened, investigating how they occurred, and what the attackers were aiming for are crucial resources to prevent security incidents from happening again. In addition, patching a security vulnerability allows developers to build more confidence in handling security issues and keeps them engaged in actively incorporating security into their software development workflow. 

\vspace{12px}

\cooltitlecomb{Psychological capabilities: }To perform security tasks, as highlighted by the COM-B model, practitioners should feel confident about executing them. Developers can gain this confidence by examining security incidents and breaches and understanding why the incidents happened in the first place. By knowing the reasons and studying a way to mitigate them, they will be more confident in assessing their own product's security. For instance,  E13 stated: ``\textit{My main outcome of being hacked is my own experience that I learned from the mistakes. I've also seen and analyzed the results of what errors other people make. So, I guess it's helping more towards my experience with security.}''

\cooltitlecomb{Reflective motivations: }Maintaining practitioners' motivations towards adopting software security practices is a complicated task. However, learning about security incidents allows developers to reflect on their practices and keep them motivated. In addition, they become aware that the chances of having a security breach are not small if they do not incorporate security into their product. For example, E7 pointed out: ``\textit{The hack of other organizations was an alert; we need to make sure that we do it better than whatever they did. And that's when we said let's introduce security restrictions to everything to maximum essentially. So there were many security features available to us that we were not using because they were potentially cumbersome.}''

\cooltitle{D14: Fear of non-adoption consequences.}
Practitioners considered that acknowledging the negative consequences of not adopting security in software development influences their need for adopting security.

\cooltitlecomb{Psychological capabilities: }Awareness of the dire consequences of not adopting security in the development workflow is highly motivating for software practitioners. Among the top three adverse effects that practitioners perceive relevant are:  losing time due to the considerable amount of time organizations need to invest in fixing vulnerabilities; losing money due to the number of resources required for patching and reinforcing security in the software product as well as in the engineering pipeline; and losing reputation which could lead to losing customers, subsequently causing severe financial issues in the organization. For example, E12 pointed out: ``\textit{The negative consequences of not adopting; it's undoubtedly a risk of breaches, data leaking, and risk of people getting access to something they shouldn't. And If you take it to the extreme, somebody could look remote in and wipe your whole system.}''

\cooltitlecomb{Reflective motivations: }The severe negative consequences of not adopting software security practices makes developers reflect on to what extent their software product is ``\textit{secure enough}'' and motivates them to minimize any possibility of exposure to exploits due to vulnerabilities in the software. For instance, E5 highlighted: ``\textit{I would like to protect my clients. I would like to make sure that I'm protected. I would like to make sure that our reputation is protected, then I'll be able to sleep at night.}''

\cooltitlecomb{Automatic motivations: }One of the main motives for practitioners to think about security in their software development workflow is to avoid experiencing a security incident. However, the fear of exposing confidential information is always a good reminder for developers to be careful about security implementations. For example, E6 emphasized: ``\textit{Adopting security is part of being a professional software developer and keeping your job's quality level and ethics. So it's in some sense, the quality of your work reflects your quality as a professional. So what would be very bad for the client would be very bad for me because I would probably get financial and legal charges and personal repercussions.}''

\cooltitle{D15: Knowledge of benefits.}
Practitioners recognized that awareness of the advantages of adopting security might indirectly influence their perspectives towards prioritizing security. For instance, most developers perceived that organizations reinforcing security practices build confidence among their employees to ensure a secure product for their customers. However, developers do not see the immediate benefits of adopting security in many situations; instead, they see the disadvantages of its adoption. 

\cooltitlecomb{Reflective motivations: }Awareness of the benefits or the importance of adopting security practices makes practitioners reflect on the value of introducing security at earlier stages of the software development pipeline. This way, they will avoid severe costs due to security bug fixing or paying ransomware. For instance, E7 stated: ``\textit{Starting with security much earlier in your design than what we see is like quick-to-market things. That reduces the chances that you will have to deal with emailing all your customers because of security leaks or paying fines for losing data or getting hacked and losing all your information due to ransomware attacks and paying that off.}''  
  
\cooltitlecomb{Automatic motivations: }Practitioners acknowledged that they felt motivated to adopt security practices because they recognized the value of security as a selling point. Organizations that care about the security and privacy of their customers' data allow organizations to operate at a bigger scale and use it for advertising and promoting trustworthiness in their products and services. For instance, E4 stated: ``\textit{Security is my organization's most important selling point. Without security, we can't operate at scale.}'' Additionally, developers indicated that they feel satisfaction when accomplishing good work by adopting security. Developers consider it rewarding to be able to prevent future headaches due to security vulnerabilities exploits. For example, E10 emphasized: ``\textit{It makes me feel like I've done a better job and that I'm helping prevent future pain from everyone else on my team.}''

\subsection{Providing contextual information to motivate developers to write secure code}
\begin{figure}
	\centering
	\includegraphics[width=0.5\textwidth]{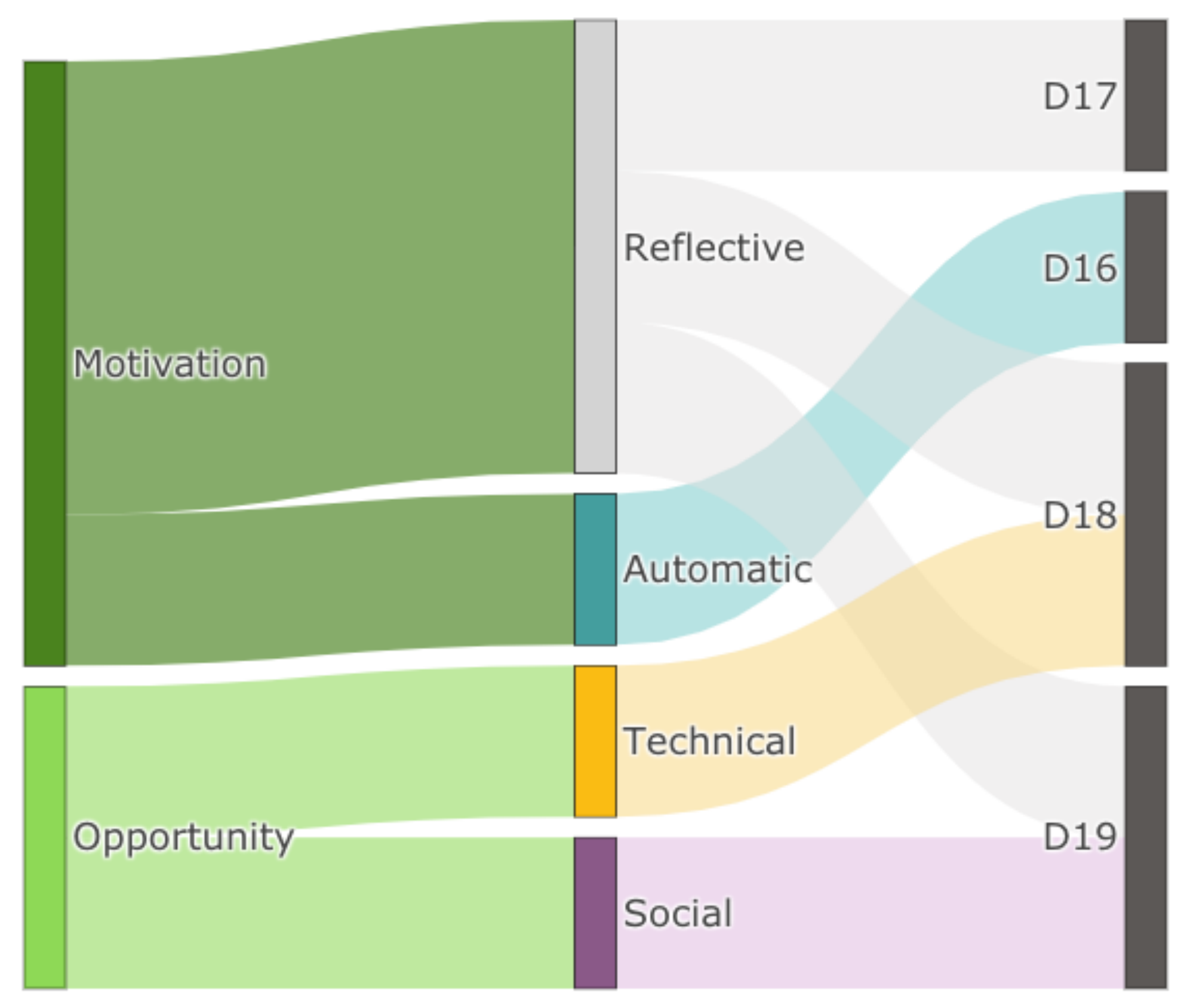}
	\caption{\textbf{Providing contextual information to motivate developers to write secure code:} 
	\textbf{D16:} Promoting a customer satisfaction/protection mindset, 
	\textbf{D17:} Awareness of the influential role of the industry type in developers' disposition towards security compliance, 
	\textbf{D18:} Awareness of developers' perceptions of the need for software security based on application characteristics, and
	\textbf{D19:} Aligning the perspective of what ``good enough'' security means.
	}
	\label{fig:category4}
\end{figure}

Practitioners recognized that being aware of the context for adopting software security matters. For example, contextual information such as industry type, application characteristics, and the importance of customers' data significantly influence developers' perceptions of the need to adopt software security practices. Figure~\ref{fig:category4} shows the relationships between the COM-B model components and drivers.

\cooltitle{D16: Promoting a customer satisfaction/protection mindset.}
Organizations promoting a software security mindset oriented towards protecting customers' data better communicate the need to adopt software security practices. Developers pointed out that feeling responsible for protecting customers' data influences their mindset and priorities for security. 

\cooltitlecomb{Automatic motivations: }Practitioners feel satisfaction from adopting security practices that ensure a secure and reliable product for their customers. They feel it is their responsibility to protect customers' private information from being exposed to malicious hackers. This is particularly relevant in the case of safety-critical systems such as health care applications. For example, E15 emphasized: ``\textit{My motivation is that people should be confident that their doctor and their medical record are in good hands and that somebody does not steal it and keeps being private. So health care can continue providing services without the system being down because of some hacker attack.}''    

\cooltitle{D17: Awareness of the influential role of the industry type in developers' disposition towards security compliance.}
Practitioners noticed how the prioritization of security significantly differs among some industries. For example, if developers are not dealing with sensitive information, as some practitioners disclosed in the gaming industry, they might think spending resources and time on security is unnecessary.

\cooltitlecomb{Reflective motivations: }In contrast to the gaming industry, some industries are more known for their security requirements. For instance, developers that work with sensitive data such as customers' payment information are more likely to feel the need to pay attention to the security implications of their technical decisions. Therefore, they feel motivated to adopt software security practices. For example, E9 emphasized: ``\textit{Major customers are coming from the financial services, so my security concern is very high.}'' Additionally, E16 pointed out: ``\textit{In the video game industry, software security is not taken seriously. Some industries require like PCI compliance but not for video game industry.}''  

\cooltitle{D18: Awareness of developers' perceptions of the need for software security based on application characteristics.}
An application's characteristics remarkably influence practitioners' perception regarding the need to adopt software security practices. Developers perceived that the security considerations in their technical decisions differ significantly depending on features such as the application type, volume of users, or to what extent applications are exposed to internet traffic.   

\cooltitlecomb{Technical opportunities: }Practitioners indicated that not all applications have the same security concerns. The application's technical specifications play a crucial role in identifying the value of introducing security practices. In addition, security requirements might scale to a different level in the system architecture, i.e., at a network or infrastructure level instead of at an application level. For example, E14 highlighted: ``\textit{Because usually, my code or application uses something that doesn't need input. If you don't have an input, usually, it works by itself. So what you need to check is that your platform or your server is secure, but not your code because there's nothing outside that can change it.}''       

\cooltitlecomb{Reflective motivations: }Web applications, mobile applications, and embedded systems have different characteristics, and therefore, multiple security considerations. Developers do not feel motivated to adopt security when developing specific applications due to their perception of the reduced likelihood of exposure to potential risks. For instance, developers who write code for embedded systems feel security is not a significant issue while developing software, or they are not even aware of security guidelines for this particular type of application. For example, E17 pointed out: ``\textit{Security is related to the type of your application. We develop embedded systems, so I don't know if there are security rules. Thus, we do not care about the security of our code. It should be secure from a higher level.}''     

\cooltitle{D19: Aligning the perspective of what ``good enough'' security means.}
Practitioners had a negative perception of the conflicts among different stakeholders involved in security decisions. Each stakeholder has a different priority and perspective for security, and these differing views about how the organization should apply security in software development become barriers to its adoption. Developers considered it critical for organizations to align stakeholders' perspectives to mitigate this barrier.    

\cooltitlecomb{Social opportunities: }Practitioners indicated that when several stakeholders are involved in security decisions, conflicts naturally arise due to the different perspectives and priorities each stakeholder holds regarding security. For example, E9 highlighted: ``\textit{POs usually get together when the definition of done includes security concerns. POs and the security team should clarify and check them. The stakeholders would be mainly the PO, the user, and the security team in this case. Discussions are about understanding why the bug is an issue, how to fix it, and whether to prioritize it.}''

\cooltitlecomb{Reflective motivations: }Developers perceived that believing they have a fully secure software product is not realistic; there is always something to improve or learn about software security. This uncertainty keeps them motivated to stay up-to-date and adopt security practices to minimize potential risks. For instance, E5 stated: ``\textit{I believe we're doing a reasonable job, but I would like to continue improving it. So, we use things that we know are very high risk, such as credit card processing. We use third-party services which are certified to handle all of that sort of stuff. But we believe our software is secure, but I know that we could do better. I know that things are stepping up regarding what's happening with the tech, and we could continue to improve substantially.}''      

\subsection{Providing justification for necessary tools and process constraints}
\begin{figure}
	\centering
	\includegraphics[width=0.5\textwidth]{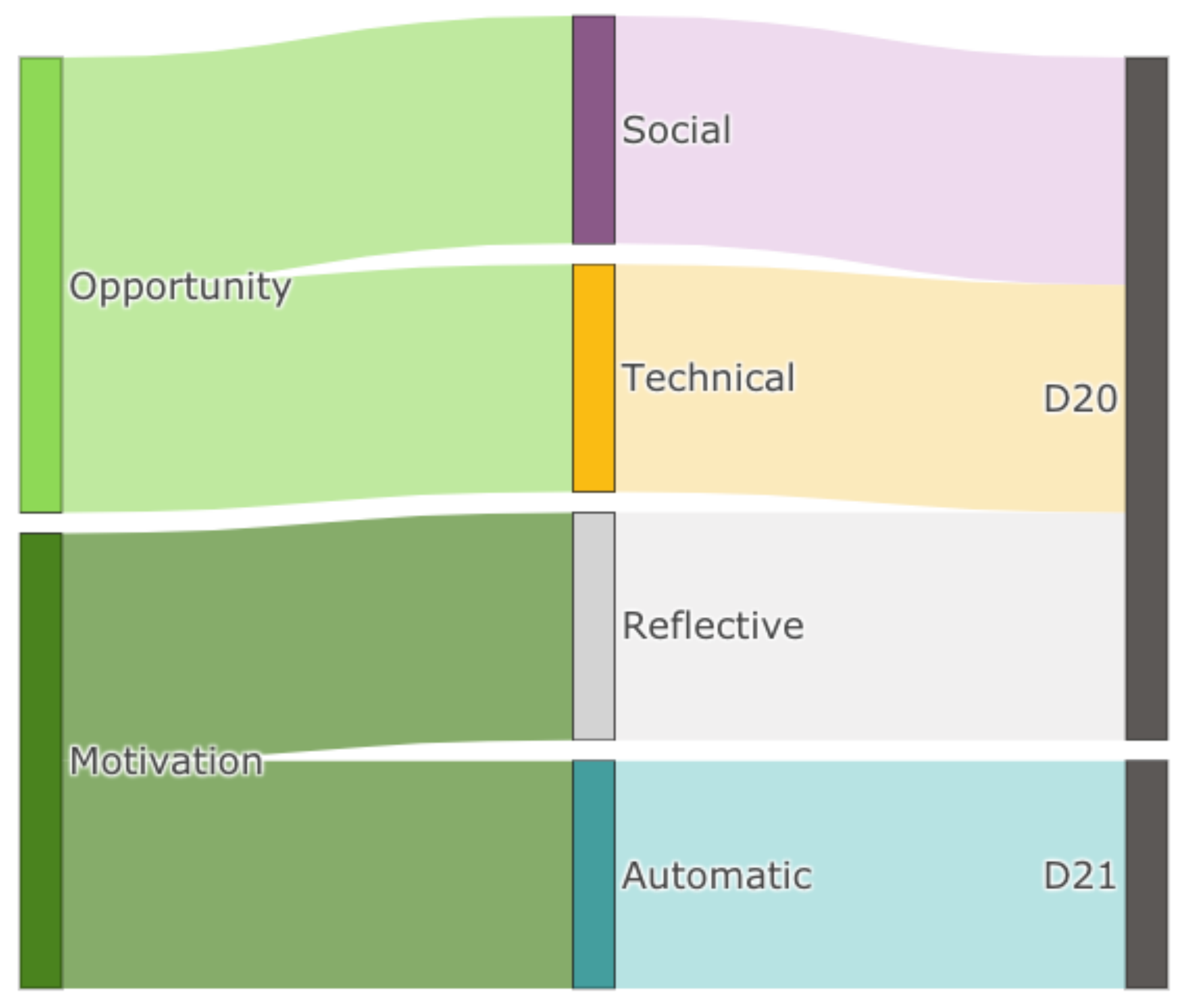}
	\caption{\textbf{Providing justification for necessary tools and process constraints:} 
	\textbf{D20:} Consideration of tool constraints on developers' autonomy, and
	\textbf{D21:} Awareness of developers' perception of security-imposed restrictions. 
	}
	\label{fig:category5}
\end{figure}

Practitioners' negative attitudes towards security are usually driven by their beliefs that adopting security restricts their freedom of choice or autonomy for selecting the most convenient tools, libraries, or technologies to perform development tasks. Figure~\ref{fig:category5} shows the relationships between the COM-B model components and drivers.  

\cooltitle{D20: Consideration of tool constraints on developers' autonomy.}
Typically, organizations with an innate security culture reinforce security practices by limiting the set of development tools, third-party libraries, or in general, any software component that developers can use in their regular workflow. Engineers perceive this situation affects their autonomy and freedom to choose the most convenient technologies to perform their work. Practitioners indicated that proper dissemination of the benefits and rationale behind those restrictions are crucial to mitigate any negative attitude caused by imposed restrictions.             

\cooltitlecomb{Social opportunities: }Some practitioners had a negative perception of the standard security practices in organizations that impose restrictive policies for using tools or any third-party component in a software project. In this context, the role of a security or compliance team is to inspect and approve any potential artifact developers may want to introduce in the project and has not been validated previously. So, for instance, E12 pointed out: ``\textit{An independent security and compliance team that has to review pretty much anything. Any external dependency you introduce has to go through the security team. Any new artifact that we produce, I think, has to be reviewed by security. So there are various checks to make sure that you aren't shipping something that could be a vector for problems.}''

\cooltitlecomb{Technical opportunities: }Some practitioners have a negative perception of adopting software security practices due to the strict restrictions while choosing any tool to develop software. For instance, E9 emphasized: ``\textit{Adopting security limits the freedom to choose what libraries we can use for development, or many security checks need to be passed.}''

\cooltitlecomb{Reflective motivations: }Developers like the freedom to choose the most suitable tool for doing their work. Organizations imposing strict restrictions negatively affect developers'  motivations to adopt security practices. For example, E13 stated: ``\textit{I think developers just like freedom of what they are doing because some people choose small companies and startups specifically for the freedom of doing what you want and how you want to do it. In big companies, it is always more restrictive.}''

\cooltitle{D21: Awareness of developers' perception of security-imposed restrictions.}
Developers perceived that adopting software security practices introduces many disruptions to their regular workflow. For instance, delays due to security inspections, waiting for feedback during security code review, or delays caused by the security team regarding the authorization to use a third-party component or library. On top of that, following security guidelines might introduce more complexity to the application, making its maintenance difficult. Therefore, software practitioners encourage organizations to identify developers' negative attitudes towards software security to foster its adoption. 

\cooltitlecomb{Automatic motivations: }Developers' motivations to adopt security are influenced by the perception that some security guidelines are unhelpful and unnecessary, especially when they introduce more complexity to the system design instead of a straightforward solution; Therefore, causing extra work in terms of maintainability. For instance, E3 emphasized: ``\textit{I work in a private network in this telecommunication company, so we are not open to the Internet. So using these Web services instead of a simple protocol, which is faster, doesn't make sense. So I think that was unnecessary. That's a barrier I've found; some security guidelines are unhelpful and unnecessary.}''

\subsection{Providing (cognitive) support to developers for writing secure code}
\begin{figure}
	\centering
	\includegraphics[width=0.5\textwidth]{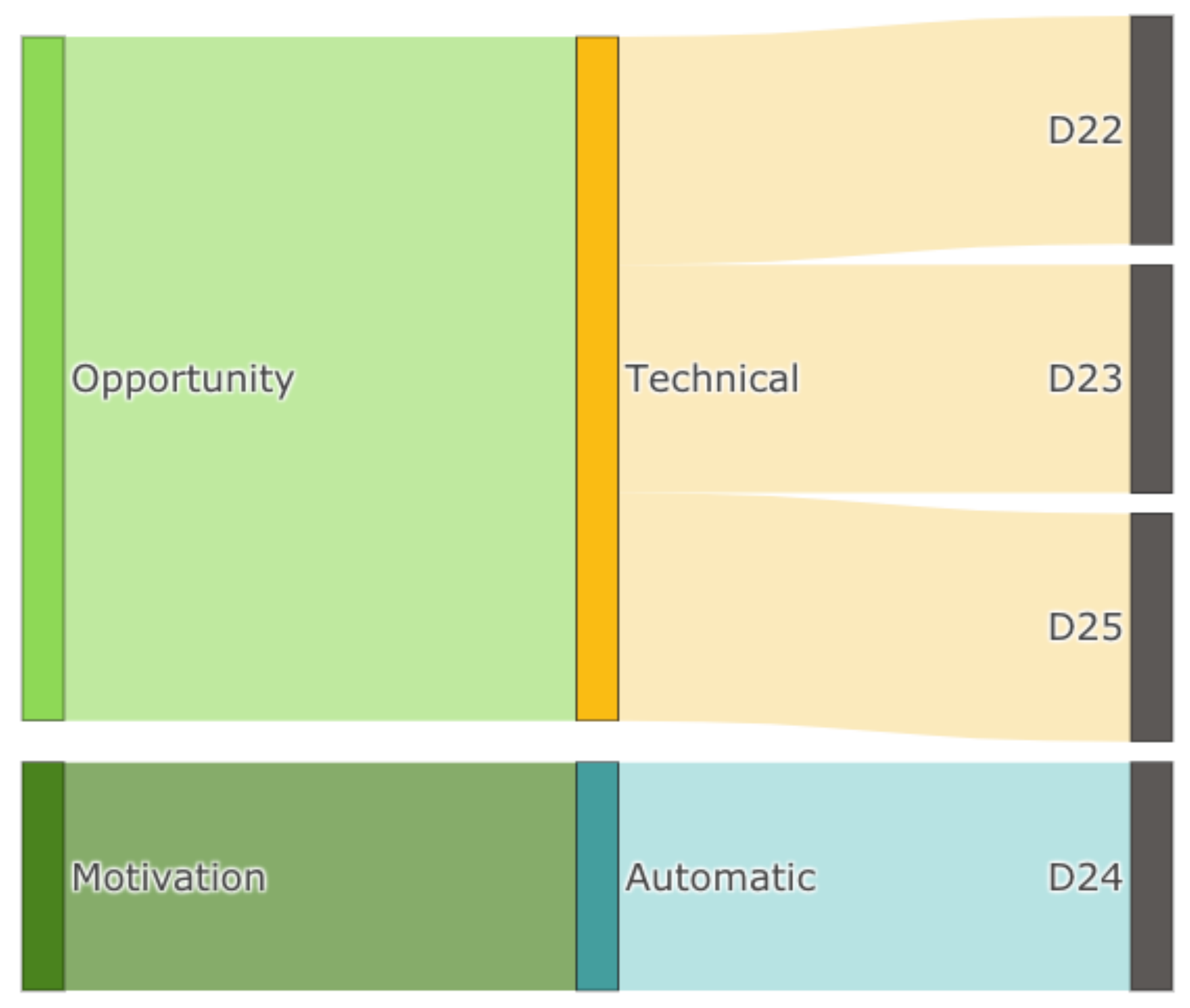}
	\caption{\textbf{Providing (cognitive) support to developers for writing secure code:} 
		\textbf{D22:} Availability of reminders, i.e., checklists, dashboards, etc.,  
		\textbf{D23:} Improving the usability (complexity reduction) and accuracy of security tools, 
		\textbf{D24:} Reducing the effort required to learn or apply security, and
		\textbf{D25:} Integrating tools into the development workflow. 
	}
	\label{fig:category6}
\end{figure}

Practitioners find it challenging to incorporate software security practices into their software development workflow due to the overwhelming number of topics they need to assimilate to write proper secure code. On top of that, developers perceive that security tools are pretty complex and sometimes inaccurate, which considerably affects their adoption. Figure~\ref{fig:category6} shows the relationships between the COM-B model components and drivers.

\cooltitle{D22: Availability of reminders, i.e., checklists, dashboards, etc.}
Most engineers indicated that software security is not a topic off the top of their heads. Reminders such as checklists are helpful resources to prevent overlooking any security considerations during code reviews. Another useful reminder perceived by developers is security tool notifications containing actionable feedback. In some particular cases, organizations with strict security policies will not allow practitioners to move forward in the development pipeline until any security concern raised by the tool is fixed.       

\cooltitlecomb{Technical opportunities: }Practitioners acknowledged that reminders are helpful to prevent overlooking security concerns while developing software. Besides security checklists, practitioners recognize the usefulness of dashboards, a visual tool that usually contains graphics and comprehensive summaries of crucial information regarding Q\&A and security metrics. For instance, E2 pointed out: ``\textit{We have a reminder to check the dashboard. Otherwise, we will forget because it's not always on top of our minds. The dashboard highlights vulnerabilities at the level of hosts, containers, and packages. In addition, it shows how many high/medium/low vulnerability issues exist in the application.}''

\cooltitle{D23: Improving the usability (complexity reduction) and accuracy of security tools.}
Practitioners pointed out that most security tools contain usability issues, making them complex to use and manage, a situation that discourages them from adopting software security practices. Additionally, security tools require sophisticated configuration to avoid a high volume of false-positive results. Default configurations result in a high level of inaccuracy and are a detriment to the value and usefulness of the tool. In this regard, developers highly appreciate security specialists' support for appropriate tools configurations.      

\cooltitlecomb{Technical opportunities: }Engineers are very conscious of the importance of security tools in adopting security practices. They recognized that better tools and security analyzers would help in their application of security practices during software development. However, some barriers to adopting security tools rely on usability and accuracy characteristics. For example, E13 highlighted: ``\textit{What would help me apply security practices? Having better tools, for sure, so better analyzers that would prompt right away where there is a security vulnerability. However, this needs to be fixed ... working here made me understand that these tools are unreliable and can give errors and false positives.}''

\vspace{10px}
\cooltitle{D24: Reducing the effort required to learn or apply security.}
Software security is a broad topic for developers. Considerable effort and dedication are required to learn and gain enough experience to write secure code correctly. Developers highlighted that organizations' efforts to reduce the scope of learning topics and provide a learning roadmap customized to developers' information needs would facilitate their disposition to adopt software security. For instance, considering topics focused on their particular applications' attack surface, programming language, frameworks, etc., are highly valued by developers.

\cooltitlecomb{Automatic motivations: }Engineers perceived that adopting software security practices demands a lot of cognitive effort, such as investing a significant amount of time for learning different techniques, absorbing a lot of information, and keeping up-to-date knowledge. These challenges highly influence developers' motivations to adopt security practices and become a significant concern for organizations interested in promoting security practices in their organizational culture. For instance, E6 stated:  ``\textit{Security shouldn't be complex. If it's difficult, people will not follow it. So it seems that it's in the company's best interest to make it straightforward for developers by giving developers clear guidelines, being more proactive, and making those guidelines available when people are dubious about if something is secure or not. So let's say a good company doesn't make security difficult.}''

\cooltitle{D25: Integrating tools into the development workflow}
Automating software security assessments is perceived by developers as a significant facilitator to adopt security practices. Furthermore, integrating security tools into the Continuous Integration (CI) pipeline allows developers to count on proper feedback regarding any potential security flaw in their source code and react accordingly. 
 
 \cooltitlecomb{Technical opportunities: }Security specialists highlighted the importance of automating security by integrating security tools into the CI/CD pipeline. This way, security scanning can be performed automatically and expose potential vulnerabilities before deployment. For example, S1 emphasized: ``\textit{Once you integrate security into the CI/CD pipeline, you make sure that your code is clean. So if you use tools like OWASP Zap~\footnote{https://www.zaproxy.org/} or Arcane, you know they're going to tell you whether you have a broken URL or you are managing passwords or secret information incorrectly, and this is going to spew them out during scanning time. Then, you can quickly fix them before you go to deployment.}''

 \subsection{Facilitating developers' acquisition of security-specific skills}
\begin{figure*}
	\centering
	\includegraphics[width=0.6\textwidth]{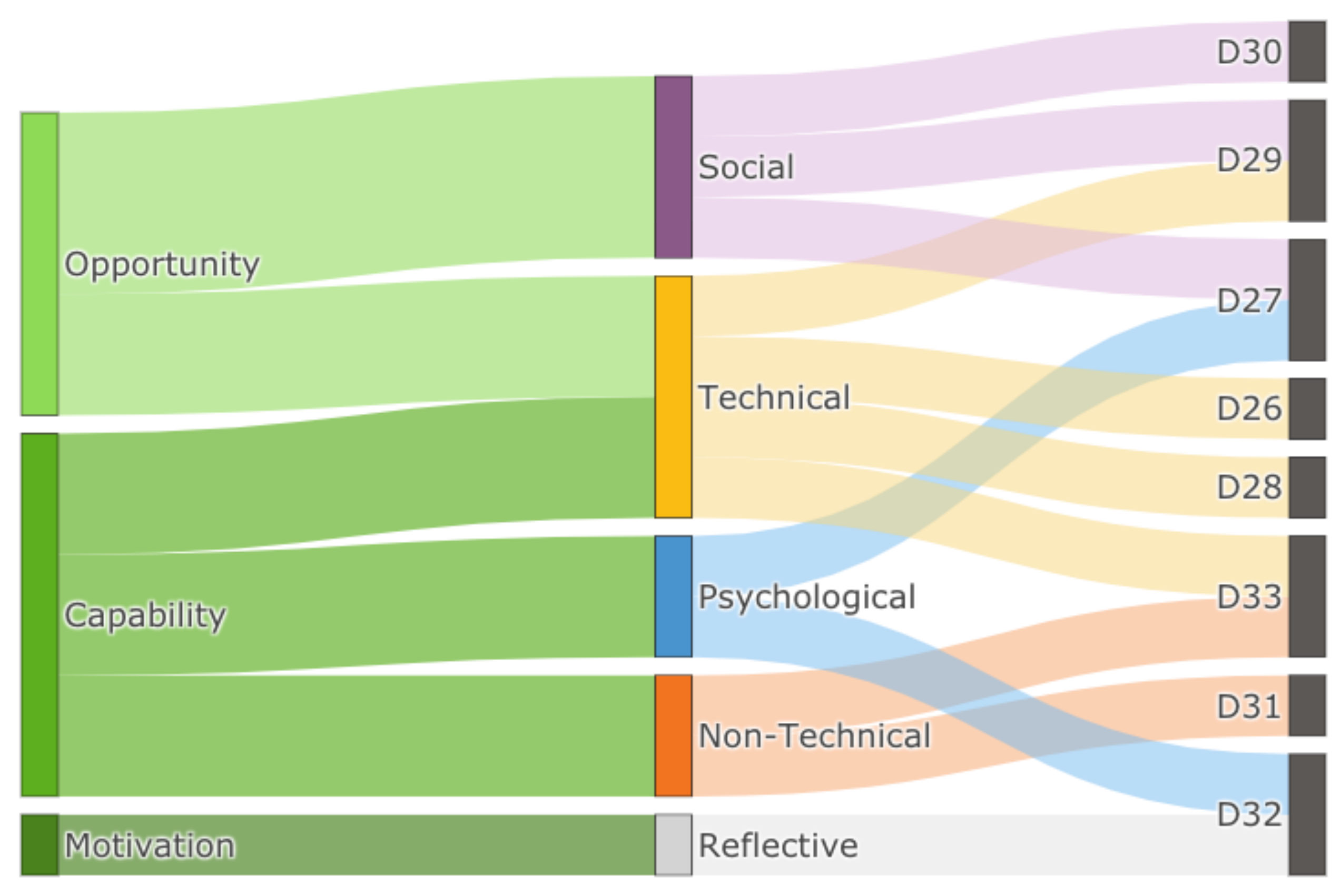}
	\caption{\textbf{Facilitating developers' acquisition of security-specific skills:} 
	\textbf{D26:} Access to learning resources, 
	\textbf{D27:} Using security practices as learning tools, 
	\textbf{D28:} Providing security education, 
	\textbf{D29:} Fostering hands-on learning / self-learning / osmosis, 
	\textbf{D30:} Creating and participating in Communities of Practice, 
	\textbf{D31:} Having non-technical skills,  
	\textbf{D32:} Confidence in their technical abilities, and
	\textbf{D33:} Awareness of necessary security skills.
	}
	\label{fig:category7}
\end{figure*}

The ability to perform security-related tasks does not necessarily depend only on technical skills. Practitioners perceived that having support from their organizations helps them acquire the necessary technical and non-technical skills to understand security challenges. In addition, support from organizations relies on providing developers with the right tools and learning resources to facilitate the adoption of software security practices. Figure~\ref{fig:category7} shows the relationships between the COM-B model components and drivers. 

 \cooltitle{D26: Accessibility to learning resources.}
Practitioners perceived that, now more than ever, there is a massive diversity of learning resources to boost technical skills in software security. For example, developers might access free online courses, conferences, specialized blogs, and publicly available security guidelines to keep up-to-date in their knowledge of security topics. 
 
\cooltitlecomb{Technical opportunities: }Practitioners indicated that they can boost their technical skills in security by benefiting from learning resources provided by organizations. An organization that offers its developers courses, training, and conferences fosters an environment for security learning. For instance, S2 stated: ``\textit{There are great webinars by OWASP. Companies like SecureIdeas, TrustedSec, Black Hills Information Security, among others, provide these worthy webinars. It's like baking cyber security into your development process because they're just like introductory level. For example, organizations, having their development team watching one of those webinars every quarter or doing a secure coding tournament from Secure Code Warrior will get developers thinking about security and start them on that path.}''

  \cooltitle{D27: Using security practices as learning tools.}
  Engineers recognized the significant value of being involved in security practices within the organization. This situation provides an effective way to learn software security. For example, developers highlighted that the feedback they receive from security code reviews, especially when the security team is involved, helps them identify anti-patterns in their coding practices and comprehend security flaws. 
  
  \cooltitlecomb{Psychological capabilities: }Engineers perceived that they gain more confidence by being involved in security practices. Experiencing how to tackle a real security issue helps developers build confidence in their capabilities. Additionally, security specialists highlighted that security games are an effective learning tool for getting developers familiar with potential threats and building a security mindset. For instance, S3 stated: ``\textit{I usually recommend STRIDE and Elevation of Privileges Game which contains 12 threats per category. It's kind of a game where developers play cards to see which threat hits on your application or your service. This game is something that they can start up with because that's around 50 threats.}''

  \cooltitlecomb{Social opportunities: }Organizations play an essential role in providing developers with experiential learning resources using security practices. Practices such as code review and pair programming help practitioners have a deeper understanding of security issues. In addition, organizations that foster knowledge sharing using security practices bring more opportunities for their developers to adopt security. For instance, E12 pointed out: ``\textit{I learned more through colleagues, best practices, and being mentored. In that regard, code review is an excellent avenue for learning. As I grew as a professional, I learned a lot through people, like pointing things out in code review.}''

 \cooltitle{D28: Providing security education.}
Engineers perceived continuous learning and proper training as crucial to mastering software security practices. Developers emphasized the vital role of organizations in providing developers with the opportunities to keep up-to-date in their security knowledge. Additionally, practitioners highlighted the need to improve security education at the university level by integrating coding best practices with security compliance source code. 

\cooltitlecomb{Technical opportunities: }Most practitioners do not take exclusively security-related courses as part of their formal education in computer science. Instead, software security is a topic that they pick up along the way while developing technical expertise. For example, E10 highlighted: ``\textit{Security was a topic included in many of those courses that had things to do with networks and systems. So there were some security topics when we got into things like developing Web applications. But I would say it was rudimentary at best. It was mostly just elementary stuff. There was much more emphasis on it when we were doing things like setting up networks or configuring a Web server on Linux or things like that. But when it comes to writing secure software, I would say that education at the university level didn't treat security as important as those other security aspects.}''

 \vspace{12px}
 
 \cooltitle{D29: Fostering hands-on learning/self-learning/osmosis.}
Practitioners recognized the value of organizations in providing developers with opportunities to learn software security by being indirectly involved in fixing security vulnerabilities or shadowing specialists while patching a security flaw. Additionally, organizations providing developers with the necessary time to learn software security by themselves or the opportunity of being mentored by a specialist are perceived by developers as highly effective methods to motivate developers to write secure code.     
  
 \cooltitlecomb{Technical capabilities: }Practitioners perceived that the ability to self-learn is crucial for starting a career in software security. A self-directed learning approach is vital to engage with the security learning process. Software security is a broad topic and requires, besides formal training, to keep up-to-date knowledge of the security exploits happening every day. For instance, S6 pointed out: ``\textit{First, I was just learning security by myself and studying on the Internet. Then I got more formal training and got involved with the community by going to conferences and giving some presentations. I think networking is a big thing in security. To start learning, I would say start small. Start with one type of vulnerability. Understand it properly, understand how to fix it. Look for that in your code or the codebase you're responsible for, and then go to other categories. We can't follow all the vulnerabilities and issues that happen every day and every week. So start small, don't try to learn about everything.}''

 \cooltitlecomb{Social opportunities: }Practitioners recognized that advocating for security within the organization is vital to demonstrate the need to adopt security practices to all  stakeholders. A starting point could be using a proof of concept to expose the security issues identified in the codebase and the implications for the organization if we overlook them. For example, S9 highlighted: ``\textit{To keep developers' motivations to adopt security, it's important to advocate for additional security controls in your environment; you can demonstrate using a proof of concept that there are severe security issues. That can help motivate folks to take action and find time to work on it. Self-learning is essential, like OWASP, and we also have the purple folks. Those are our two best resources right now, and they just partnered together to offer classes and stuff, which is fantastic. It's kind of like demystifying security in the company.}''

 \cooltitle{D30: Creating and participating in communities of practice.}
Practitioners emphasized the crucial role of communities of practice to learn software security. Practitioners perceived that a relevant characteristic of a good security culture is organizations creating and promoting an environment where people interested in security-related topics gather together to learn and share knowledge. Additionally, practitioners considered it essential to join efforts with external communities since being hacked could affect any organization or business. 
  
\cooltitlecomb{Social opportunities: }Currently, developers are using external communities to get help on security issues. For example, communities such as OWASP provide an excellent environment for practitioners to learn about security. Similarly, developers perceived that organizations that form communities inside the company and gather all the practitioners who show security interest help build a supportive environment to facilitate security learning. For instance, S8 emphasized: ``\textit{There are many gangs in the community and people doing similar things helping each other solve security issues and learning.}''

 \cooltitle{D31: Having non-technical skills.}
Besides having proper technical skills in software security, developers emphasized having non-technical or soft skills. Ensuring a secure software product is a collaborative effort and demands the participation of several stakeholders. Inevitably, conflicts between stakeholders may arise due to different priorities and perspectives of approaching and applying security. In these situations, developers emphasized the critical role of having soft skills such as communication, critical thinking, empathy, etc., to facilitate discussions and mitigate any potential conflict. 
  
\cooltitlecomb{Non-Technical capabilities: }Practitioners indicated that communication is one of the most important non-technical skills required to conduct software security practices. For instance, they believe communication is essential when describing the importance of security to other stakeholders, discussing security issues with developers, and handling security incidents. Furthermore, developers need to elaborate the situation to the management team and other stakeholders to discuss further actions when an incident happens. For example, E8 pointed out: ``\textit{Besides communication, researching is also a critical non-technical skill because you have to deeply investigate what went wrong when you are facing a security issue.}''

 \cooltitle{D32: Confidence in their technical abilities.}
 Engineers indicated that confidence in their technical abilities in software security significantly influences the adoption of software security practices. Developers' low confidence in their technical skills will discourage them from adopting security practices. Contrarily, over-confidence is perceived negatively and as a deterrent to their motivation to keep up-to-date knowledge to improve their security practices. 
  
 \cooltitlecomb{Psychological capabilities: }Practitioners perceived that confidence is essential to perform any engineering task. Developers become confident in their technical abilities to conduct software security tasks because of their experience facing security threats or due to the collective expertise of their software teams. Having someone on the team who has security experience boosts the entire team's confidence. For instance, E16 pointed out: ``\textit{Academic knowledge can get you close to ensuring secure software. However, having expertise or having someone on your team who has experienced a security threat will give you more confidence to say you have a secure product.}''

 \cooltitlecomb{Reflective motivations: }Most engineers recognized that security is not their primary focus at work; It is not a topic that they need to apply daily. However, developers are familiar with self-learning and hands-on learning techniques within a software development context, which are also essential and commonly used in software security. Therefore, it is not a drastic transition for a developer to feel motivated to dig deeper in software security and become a security champion or an application security specialist. Due to the deluge of available security information, it is also vital for practitioners to recognize that security is a never-ending learning cycle; There are new sophisticated mechanisms, tools, guidelines, and best practices to learn. Such intense and dynamic context motivates developers to aggregate security to their professional profile. For instance, E5 stated: ``\textit{So the first question is how comfortable am I with security? I think I know enough that I don't know much, but I'm not starting from zero. I know that there are many more sophisticated mechanisms out there right now. And I know that we've got a lot of development experience that we can just use it.}''

 \cooltitle{D33: Awareness of necessary security skills.}        
 Engineers perceived that awareness of what they should know to properly adopt and apply software security practices significantly influences their motivation to adopt security practices. For instance, developers considered it relevant to know about malicious hackers' mindsets, their methods, most common attack vectors, and the best coding practices to prevent them.       

 \cooltitlecomb{Technical capabilities: }Developers acknowledged that it is essential to have a solid understanding of the security basics to understand technical aspects of security exploits, follow security guidelines, and apply security patches, and therefore, prevent harmful consequences in a potential system attack. For instance, E11 highlighted: ``\textit{Knowledge of defensive programming patterns and how to sanitize inputs is valuable. Knowing about security issues and what is needed to solve them it's part of the job.}'' Additionally, E15 emphasized: ``\textit{You need to know about software architecture, software testing, many technical issues like how you design software to be secure? And you need to know what type of threats are there? How can you do the testing? How do you do penetration testing? Also, to let other people do that. So you should always think you should always question your security knowledge because you can think about everything upfront. However, still, somebody else will be able to penetrate the system.}''      
 
 \cooltitlecomb{Non-Technical capabilities: }Practitioners recognized that to ensure a secure software product, it is crucial to consider the human aspects of the product in addition to the technical features. In this context, it is valuable to acquire knowledge about the users, where the product is used, and what abnormal product usage could potentially occur. In other words, to analyze upfront what can go wrong due to a misuse of the application by the end-user. For example, E15 emphasized: ``\textit{Regarding non-technical skills, I need to have domain knowledge and knowledge about end-users. That's very important where the system is used, what can happen if that something doesn't work or goes wrong or how people enter the system, if they share their password data or how people work with that.}''


\section{Related work}
\label{sec:relatedwork}



\begin{table*}

\caption{Key drivers associated with the adoption of software security practices that have emerged in the previous literature.}

\begin{center}
	
\resizebox{1.0\textwidth}{!}{%
	\begin{tabular}{llcccccccccccccccr}
	\toprule

	&& \rotatebox{90}{\textbf{Rauf et al.~\cite{rauf2022}}} & \rotatebox{90}{\textbf{Haney et al.~\cite{haney2021a}}} & \rotatebox{90}{\textbf{Jones and  Rastogi~\cite{jones2004}}} & \rotatebox{90}{\textbf{Werlinger et al.~\cite{werlinger2009}}} & \rotatebox{90}{\textbf{Mokhberi and Beznosov~\cite{mokhberi2021}}} & \rotatebox{90}{\textbf{Lopez et al.~\cite{lopez2019}}} & \rotatebox{90}{\textbf{Xie et al.~\cite{xie2011}}} & \rotatebox{90}{\textbf{Poller et al.~\cite{poller2017}}} & \rotatebox{90}{\textbf{Sajwan et al.~\cite{sajwan2021}}} & \rotatebox{90}{\textbf{Assal and Chiasson~\cite{assal2019}}} & \rotatebox{90}{\textbf{Assal and Chiasson~\cite{assal2018}}} & \rotatebox{90}{\textbf{Votipka et al.~\cite{votipka2020}}} & \rotatebox{90}{\textbf{Weir et al.~\cite{weir2021}}} & \rotatebox{90}{\textbf{Braz et al.~\cite{braz2022}}} & \rotatebox{90}{\textbf{Fischer and Grossklags~\cite{fischer2022}}} & \rotatebox{90}{\textbf{Totals}} \\
	\midrule

	\multicolumn{16}{l}{\textbf{Building an organizational security culture}} \\
\midrule

	\hspace{1mm}\textbf{D1}&Organization promoting/mandating security & \here &  & \here &  & \here &  &  & \here &  & \here & \here &   &  & & & \textbf{6} \\
	\hspace{1mm}\textbf{D2}&Prioritizing security practices& \here &  & \here & \here & \here &  & \here & \here & \here & \here & \here &  & & \here & & \textbf{10} \\
	\hspace{1mm}\textbf{D3}&Having a security-specific role filled  &  & \here & \here &  & \here &  & \here & \here &  & \here &  & \here &  & & & \textbf{7} \\
	\hspace{1mm}\textbf{D4}&Overcoming the resistance to change &  &  & \here &  & \here &  &  & \here & \here & \here & \here &  &  & & & \textbf{6} \\
	\hspace{1mm}\textbf{D5}&Fostering collaboration between engineering and security teams  & \here &  & \here & \here & \here &  &  &  &  &  &  &  &  & & & \textbf{4} \\
	\hspace{1mm}\textbf{D6}&\thead[l]{Awareness of the social perception of security adoption in one's own\\ organization and professional network}  & \here & \here & \here &  & \here &  &  &  &  & \here & \here &  & \here & & & \textbf{7} \\
	\hspace{1mm}\textbf{D7}&Providing awareness of external incentives and compliance  & \here &  & \here &  & \here &  & \here &  & \here &  &  &  &  & & & \textbf{5} \\

	\midrule
	\multicolumn{16}{l}{\textbf{Facilitating the adoption of software security by developers}} \\
	\midrule

	\hspace{1mm}\textbf{D8}&Shaping developer's attitudes towards security  & \here &  & \here &  & \here & \here & \here & \here & \here & \here & \here & \here &  & & & \textbf{10} \\
	\hspace{1mm}\textbf{D9}&Tool awareness  & \here &  & \here &  & \here &  &  &  &  & \here & \here & \here &  & & & \textbf{6} \\
	\hspace{1mm}\textbf{D10}&Standard guidelines geared at developers  & \here &  & \here &  & \here &  &  &  &  &  &  &  &  & & & \textbf{3} \\
	\hspace{1mm}\textbf{D11}&Reduction of system complexity  & \here &  &  &  & \here &  & \here &  &  &  &  & \here &  & & & \textbf{4} \\
			
	\midrule
	\multicolumn{16}{l}{\textbf{Understanding risks, benefits, and trade-offs}} \\
	\midrule
	
	\hspace{1mm}\textbf{D12}&Awareness of potential risks and security incidents  & \here &  & \here & \here & \here & \here &  & \here & \here & \here &  &  &  & \here & \here & \textbf{10} \\
	\hspace{1mm}\textbf{D13}&Learning from actual incidents  &  &  &  &  & \here &  &  &  &  &  &  &  &  & & & \textbf{1} \\
	\hspace{1mm}\textbf{D14}&Fear of non-adoption consequences  &  &  &  &  & \here &  &  &  &  & \here & \here &  &  & & & \textbf{3} \\
	\hspace{1mm}\textbf{D15}&Knowledge of benefits  &  &  & \here &  & \here &  &  &  &  & \here &  &  & \here & & & \textbf{4} \\

\midrule
	\multicolumn{16}{l}{\textbf{Providing contextual information to motivate developers to write secure code}} \\
	\midrule
	
	\hspace{1mm}\textbf{D16}&Promoting a customer satisfaction/protection mindset  &  & \here &  &  & \here &  & \here & \here & \here & \here & \here &  &  & & & \textbf{7} \\
	\hspace{1mm}\textbf{D17}&\thead[l]{Awareness of the influential role of the industry type in developers'\\disposition towards security compliance}  &  &  &  &  &  &  &  &  &  &  &  &  &  & & & \textbf{0} \\
	\hspace{1mm}\textbf{D18}&\thead[l]{Awareness of developers' perceptions of the need for software security \\based on application characteristics} &  &  &  &  &  &  & \here &  & \here & \here &  &  &  & & & \textbf{3} \\
	\hspace{1mm}\textbf{D19}&Aligning the perspective of what "good enough" security means& \here &  & \here & \here & \here &  &  & \here &  & \here &  &  &  & & & \textbf{6} \\

	\midrule
	\multicolumn{16}{l}{\textbf{Providing justification for necessary tools and process constraints}} \\
	\midrule
	
	\hspace{1mm}\textbf{D20}&Consideration of tool constraints on developers' autonomy &  &  &  &  &  &  &  &  &  &  &  &  &  & & & \textbf{0} \\
	\hspace{1mm}\textbf{D21}&Awareness of developers' perception of security-imposed restrictions &  &  & \here &  & \here &  & \here &  & \here & \here &  &  &  & & & \textbf{5} \\

	\midrule
	\multicolumn{16}{l}{\textbf{Providing (cognitive) support to developers for writing secure code}} \\

	\midrule
	
	\hspace{1mm}\textbf{D22}&Availability of reminders, i.e., checklists, dashboards, etc.& \here &  &  &  &  &  &  &  &  &  &  &  &  & \here & \here & \textbf{3} \\
	\hspace{1mm}\textbf{D23}&Improving the usability (complexity reduction) and accuracy of security tools& \here &  &  & \here & \here &  &  &  &  & \here &  &  &  & & & \textbf{4} \\
	\hspace{1mm}\textbf{D24}&Reducing the effort required to learn or apply security  & \here &  &  &  & \here &  & \here &  &  & \here &  &  &  & & \here & \textbf{5} \\
	\hspace{1mm}\textbf{D25}&Integrating tools into the development workflow  &  &  &  &  & \here &  &  &  &  &  &  &  &  & & & \textbf{1} \\			

	\midrule
	\multicolumn{16}{l}{\textbf{Facilitating developers' acquisition of security-specific skills}} \\
    \midrule
    
	\hspace{1mm}\textbf{D26}&Accessibility to learning resources  & \here & \here & \here &  & \here &  &  &  & \here &  &  & \here &  & & & \textbf{6} \\
	\hspace{1mm}\textbf{D27}&Using security practices as learning tools  &  &  &  &  &  &  &  &  &  &  &  &  &  & & & \textbf{0} \\
	\hspace{1mm}\textbf{D28}&Providing security education  &  &  & \here &  & \here &  &  & \here & \here &  &  &  & \here & & & \textbf{5} \\
	\hspace{1mm}\textbf{D29}&Fostering hands-on learning/self-learning/osmosis  &  &  &  &  & \here &  &  & \here & \here &  &  &  &  & & & \textbf{3} \\
	\hspace{1mm}\textbf{D30}&Creating and participating in communities of practice  & \here &  &  &  & \here &  &  & \here & \here &  &  &  &  & & & \textbf{4} \\
	\hspace{1mm}\textbf{D31}&Having non-technical skills  & \here & \here &  & \here & \here & \here &  &  &  &  &  &  &  & & & \textbf{5} \\
	\hspace{1mm}\textbf{D32}&Confidence in their technical abilities  & \here & \here & \here &  & \here & \here &  &  & \here & \here &  & \here &  & \here & & \textbf{9} \\
	\hspace{1mm}\textbf{D33}&Awareness of necessary security skills  & \here & \here & \here &  & \here & \here &  &  & \here & \here &  & \here & &   & & \textbf{8} \\

	\midrule
	& \textbf{Totals} & \textbf{19} & \textbf{7} & \textbf{18} & \textbf{6} & \textbf{28} & \textbf{5} & \textbf{9} & \textbf{10} & \textbf{14} & \textbf{18} & \textbf{8} & \textbf{7} & \textbf{5} & \textbf{3} & \textbf{3} &  \\

	\bottomrule
	\end{tabular}%
   }

\vspace{0.02in}

\end{center}

\label{tab:relatedwork}

\end{table*}

We briefly provide an overview of the related research and how it aligns with the drivers (shown in italics) our study revealed. 
We searched the ACM, IEEE Xplore, and Springer research databases and found 15 papers that studied the adoption of software security practices, many of which were published quite recently. Two of these papers were systematic literature reviews of work on this topic~\cite{mokhberi2021,rauf2022}. 
We found that most (30/33) of the drivers that emerged from our study also appeared in this other research. Although most papers reported between 3 and 19 of the drivers we found, Mokhberi and Beznosov's~\cite{mokhberi2021} systematic literature review reported 28 of our drivers.  
This overlap further strengthens the relevance of our work, but we were surprised that several of the drivers we found to be quite important were not mentioned in these 15 papers. These drivers are \textit{(1) awareness of the influential role of the industry type in developers' disposition towards security compliance}, \textit{(2) consideration of tool constraints on developers' autonomy}, and \textit{(3) using security practices as learning tools.} 
We summarize how our research relates to the related research in Table~\ref{tab:relatedwork}. 
This table can be used to not only see which of the drivers have been reported before, but it also provides an index into further reading about these drivers, while highlighting that some of them may perhaps call for further research. 

\cooltitle{Building an organizational security culture.}
Some researchers have extensively studied how organizations build a security culture, specifically \textit{organizations promoting or mandating security (D1)}. For instance, Rauf et al.~\cite{rauf2022}, Jones and Rastogi~\cite{jones2004}, and Mokhberi and Beznosov~\cite{mokhberi2021} exposed the lack of security culture in teams and organizations as a significant deterrent to the adoption of security. In particular, Jones and Rastogi highlighted that management should be responsible for disseminating the security policies, standards, guidelines, and procedures across all teams in the organization~\cite{jones2004}. Additionally, some research literature has revealed the relevant role of organizations in \textit{prioritizing security practices (D2)}. For example, Mokhberi and Beznosov~\cite{mokhberi2021} and Poller et al.~\cite{poller2017} agreed that organizations that do not provide the necessary resources prevent developers from implementing security. Specifically, Poller et al. pointed out that when managers see security as a resource conflict with feature development, developers also perceive implementing security as not worth the time and energy~\cite{poller2017}.                  

Several other researchers have emphasized the critical function of a \textit{security-specific role in the organization (D3)}. For example, Xie et al.~\cite{xie2011} pointed out that security experts usually act as security supervisors of the whole development process. However, Xie et al.~\cite{xie2011} also indicated in the same study that developers might exhibit a more relaxed attitude towards security when there are experts to back them up. Moreover, Poller et al.~\cite{poller2017} highlighted that security inspections conducted by external security consultants become an eye-opener, fostering awareness among developers about the security topics they need to look after in their daily work. Furthermore, some researchers recognized the importance of the \textit{awareness of the social perception of security adoption (D6)}. For example, when the whole team is responsible for security, the motivation for adopting and implementing security could have a snowball effect and lead to motivating more team members to acknowledge the value of adopting security~\cite{assal2018,assal2019}.

\cooltitle{Facilitating developers software security adoption.}
Other researchers have found that organizations play a crucial role in \textit{shaping developers' attitudes towards security (D8)}. For instance, Rauf et al.~\cite{rauf2022}, Jones and Rastogi~\cite{jones2004}, and Mokhberi and Beznosov~\cite{mokhberi2021} found that developers usually do not perceive the usefulness of security practices. Their studies highlighted that most developers might have an attitude that security is someone else's responsibility~\cite{rauf2022}, or perceive it as a hindrance~\cite{jones2004}, or in contrast, consider security to be a shared responsibility~\cite{mokhberi2021}. Furthermore, other researchers also emphasized that interaction through a gamification approach is an effective tool to engage developers in security practices as developers often enjoy the physical aspects of a game~\cite{lopez2019}.

Several researchers have pointed out \textit{tool awareness (D9)} as a relevant driver. For example, the lack of awareness of security tools and vulnerabilities~\cite{rauf2022} reduces the likelihood of developer involvement in security practices. A similar lack of adoption occurs when organizations do not provide the proper training for using security tools. As a result, developers usually use security tools without a complete understanding of tool functionality~\cite{mokhberi2021}. Additionally, a few other researchers have emphasized the importance of organizations providing \textit{standard guidelines geared at developers (D10)} and the \textit{reduction of system complexity}. For instance, Mokhberi and Beznosov~\cite{mokhberi2021} reported that developers often face a lack of general security guidelines and no one in charge of ensuring that those security requirements are followed.                

\cooltitle{Understanding risks, benefits, and stakeholders' trade-offs.}
Several researchers have recognized the value of understanding risks, benefits, and stakeholders' trade-offs, in particular, the  \textit{awareness of potential risks and security incidents (D12)}. For instance, Rauf et al.~\cite{rauf2022} pointed out that developers might misplace trust on frameworks or third-party APIs. As a result, developers can introduce vulnerabilities into the source code, assuming that frameworks or libraries properly handle security by default. Additionally, Lopez et al.~\cite{lopez2019} highlighted that public incidents enable information trading and risk awareness. Developers usually build awareness by expanding on technical information and providing additional scenarios and examples from their personal experiences.

Other researchers have emphasized the crucial role of drivers such as \textit{fear of non-adoption consequences (D14)} and \textit{knowledge of benefits (D15)}. For instance, organizations' security efforts are less effective when developers perceive a disinterest in adopting software security practices. This situation usually happens when there are no perceived negative consequences to the customers or the business from the lack of security in the SDLC~\cite{assal2018}. Additionally, Assal and Chiasson~\cite{assal2019} highlighted that developers feel motivated to adopt security practices when they are aware of similar software (to the one they work on) suffering a security breach---this situation becomes an ``eye-opener'' for them. Finally, Mokhberi and Beznososv~\cite{mokhberi2021} recognized that \textit{having experienced a real security issue (D13)} is the primary driver that increases awareness and concerns about security among developers in the long run. As a result, adopting security practices and learning about security mechanisms to protect their code becomes a priority. 

\cooltitle{Providing contextual information to motivate developers to write secure code.}
Some researchers have emphasized the relevant role of the organization in providing contextual information to motivate developers to write secure code. In particular, Xie et al.~\cite{xie2011} highlighted the importance of \textit{promoting a customer satisfaction/protection mindset (D16)}. They pointed out that a critical motivator for developers is the concerns of the customer or client: If the customer cares about security, the company has to care about security. Furthermore, Poller et al.~\cite{poller2017} confirmed a similar result, highlighting that any feedback from the customer motivates developers to write secure code. Furthermore, Assal and Chiasson~\cite{assal2019} emphasized that developers who care about their users' security and privacy feel encouraged to adopt security practices.

Other researchers have pointed out that \textit{aligning the perspective of what ``good enough'' security means (D19)} and \textit{awareness of developers' perceptions of the need for software security (D18)} are vital drivers that organizations should pay careful attention to in order to encourage developers to implement security practices. For instance, Werlinger et al.~\cite{werlinger2009} highlighted that developers have to communicate with other stakeholders that hold different perceptions of risks, sometimes considering security as a second priority and not having security culture training. Therefore, developers feel the need to persuade these stakeholders of the importance of security controls, which sometimes becomes frustrating. Additionally, Xie et al.~\cite{xie2011} highlighted that \textit{developers' perceptions of the need for software security (D18)} are influenced by the applications' characteristics. For instance, they pointed out that middleware developers might not be concerned about adopting security practices since they believe security should only be an issue for front-end applications. Moreover, Assal and Chiasson~\cite{assal2019} emphasized that developers' perceptions of the need for software security might be influenced by false assumptions that the software they develop is not prone to security attacks. They might also believe that users will not be technically capable of doing anything malicious for fear of losing their jobs. Interestingly, no study reported the importance of \textit{developers' perceptions and disposition towards security compliance based on the type of business (D17)} they develop software for.

\cooltitle{Providing justification for necessary tools and process constraints.}
Researchers have agreed that the \textit{awareness of developers' perception of security-imposed restrictions (D21)} is a crucial driver to motivate developers to write secure code. Specifically, Jones and Rastogi~\cite{jones2004} emphasized that developers perceive security as a barrier to functionality, adding constraints and reducing flexibility. Additionally, Xie et al.~\cite{xie2011} highlighted that developers consider security as an expense and potentially time-consuming activity. So when the budget is limited, software security is one of the concerns that can be overlooked. Furthermore, sometimes developers commonly perceive that by focusing more on software security, companies might lose their business opportunities~\cite{assal2019}. Surprisingly, we found no studies that reported the \textit{consideration of tool constraints on developers' autonomy (D20)}.

\cooltitle{Providing (cognitive) support to developers for writing secure code.}
Researchers have recognized that organizations should provide developers with cognitive support to facilitate the adoption of security practices. Specifically, Werlinger et al.~\cite{werlinger2009} emphasized that developers feel discouraged to adopt security practices when \textit{security tools' usability and accuracy (D23)} become part of the problem instead of being a facilitator to fix security vulnerabilities. For instance, they pointed out several security tool issues that require attention from tool providers, such as better support for collaboration, decreased complexity, support to disseminate knowledge, flexible reporting, and better integration of security tools with communication channels used in an organization. Additionally, other researchers recognized that \textit{reducing the effort required to learn or apply security (D24)} is a relevant driver that organizations should not overlook. For example, Fischer and Grossklags~\cite{fischer2022} proposed encouraging developers to write secure code by providing them with reminders and recommendations that prioritize security. In this way, developers would be capable of making safer choices that lead to writing more secure code. They performed two experiments that nudged developers while copying/pasting code from Stack overflow and searching for code snippets in Google.

A few researchers have also highlighted that it is important for organizations to \textit{integrate tools into the development workflow (D25)}~\cite{mokhberi2021} and \textit{provide developers with reminders (D22)}~\cite{rauf2022,fischer2022}. For instance, Rauf et al.~\cite{rauf2022} emphasized that developers often add security as an afterthought and forget to give attention to secure coding practices. Therefore, there is a need to remind developers about security concerns while developing software. However, Braz et al.~\cite{braz2022} highlighted that during code reviews, developers, despite receiving a tailored security checklist as a reminder, they can not find more vulnerabilities than when are just instructed to focus on security issues. Additionally, Mokhberi and Beznosov~\cite{mokhberi2021} acknowledged that a lack of integration with the development environment becomes a deterrent for developers to use security tools and reduces their engagement with security practices.  


\cooltitle{Facilitating developers' acquisition of security-specific skill sets.}
Several researchers have highlighted three vital drivers to motivate developers to write secure code: having \textit{confidence in their technical abilities  (D32)}, an \textit{awareness of the necessary security skill set  (D33)}, and \textit{accessibility to learning resources  (D26)}. For instance, Mokhberi and Beznosov~\cite{mokhberi2021} pointed out the role of personality as one of the human dimensions of developers' challenges in engineering secure software. Lack of confidence and false confidence are reasons developers mistakenly believe that their code is secure. Thus, they are unable to recognize vulnerabilities in their code. Additionally, Votipka et al.~\cite{votipka2020} acknowledged that the primary reason why teams do not implement security is due to a lack of knowledge and, above all, experience in different types of vulnerabilities. Furthermore, Sajwan et al.~\cite{sajwan2021} emphasized that organizations usually employ traditional training resources and methods that developers do not feel are practical and actionable. Most of these learning resources focus on policies and protocols, reading, watching videos, or office conversation by either internal teams or external parties.        

Other researchers have highlighted four essential drivers that influence developers in the adoption of  security practices: \textit{providing security education  (D28)}, \textit{having non-technical skills  (D31)}, \textit{creating and participating in communities of practice  (D30)}, and \textit{fostering hands-on learning, self-teaching, and osmosis  (D29)}. For instance, Weir et al.~\cite{weir2021} pointed out that security-related workshops facilitated by managers appear more effective than those facilitated by developers or security specialists. Additionally, Haney et al.~\cite{haney2021a} emphasized the relevance of having interpersonal skills, in particular, communication skills for dealing with all stakeholders involved in a security issue. Furthermore, researchers have reported that knowledge sharing is crucial for learning security practices. Developers learn the best from talking to other people in their teams as they can learn more technical skills while applying existing knowledge~\cite{sajwan2021}. Researchers have also emphasized the crucial role of peer-based learning as an effective method to learn security practices. Developers usually perceive mentoring as an effective way to understand the rationale behind threats and techniques to mitigate them~\cite{weir2021}. Interestingly, no previous studies have pointed out the essential role of \textit{using security practices as learning tools (D27)}.

\cooltitle{Comparing and contrasting drivers identified  in our study with current literature.} In the above, we compared the drivers from our study with those found in the literature we reviewed. 

In particular, the Mokhberi and Beznosov's~\cite{mokhberi2021} recent systematic literature review is the closest to our work. They presented a set of 17 areas of challenges across three dimensions: human, organizational, and technological. Their results align with 28 of our drivers, but they mentioned two factors that did not emerge from our findings. First, sometimes \textit{responsibilities and roles defined in the software teams might conflict with applying security}, and therefore, developers will not follow security practices. Second, \textit{developers may misuse APIs/libraries}, which leads to decreasing security in an application, making it easier to exploit. 
Another study related to our work is the systematic literature review conducted by Rauf et al.~\cite{rauf2022}. The authors presented a catalog of factors that influence developers' security behavior. Their work introduced 17 internal factors and 11 external factors analyzed from three different perspectives: \textit{knowledge deficit}, \textit{attention deficit}, and  \textit{intention deficit}. Their set of factors aligns with 19 of our drivers, and they did not mention any other drivers that we did not find in our study.

As we indicated before, three of our \numberofdrivers drivers have not been identified in the literature we reviewed. They are \textit{D17: awareness of the influential role of the industry type in developers' disposition towards security compliance}, \textit{D20: consideration of tool constraints on developers' autonomy}, and \textit{D27: using security practices as learning tools}. 
We were surprised that these drivers did not appear in any of the 15 papers we reviewed (two of which were quite comprehensive systematic reviews). 
In the related literature, we did not find any studies that highlighted how developers' disposition towards security compliance is highly influenced by, what the general industry mindset considers relevant, in terms of security. 
For example, developers working in the game industry or chip manufacturing tend to be more reluctant to adopt software security practices because the entire industry focuses on security, not at the application level but the infrastructure level. Additionally, researchers have given less attention to developers' perceptions regarding how the use of tools for security could affect their autonomy. Therefore, tools can become a deterrent to adopting security due to the restrictions imposed on developers' workflows. Moreover, we did not find related literature that explores the effects of security learning when performing security practices, even though the feedback collected while implementing a security practice can be a tremendous learning tool for developers. Section~\ref{sec:results} describes these drivers in more detail.

	\section{Discussion}
	\label{sec:discussion}

In this section, we review the implications of our findings.
First, we discuss the implications for organizations, and our recommendations for developers and security specialists when adopting or advocating security practices. Then, we discuss how researchers can use the power of behavioral theories to understand and frame research on software security. 

\subsection{Implications for Organizations}

Since organizations play the most crucial role in fostering the adoption of software security practices, our study pointed to several recommendations for them. In the following, we discuss how they can use the \frameworkname framework as a diagnosis tool to help identify which aspects of their security practices could be improved. Then, we discuss what organizations need to consider when understanding the perspectives of developers and security specialists, and additional considerations they should be aware of when applying the \frameworkname  framework.

\subsubsection{How to use the \frameworkname framework}
\begin{figure*}
	\centering
	\includegraphics[width=0.8\textwidth]{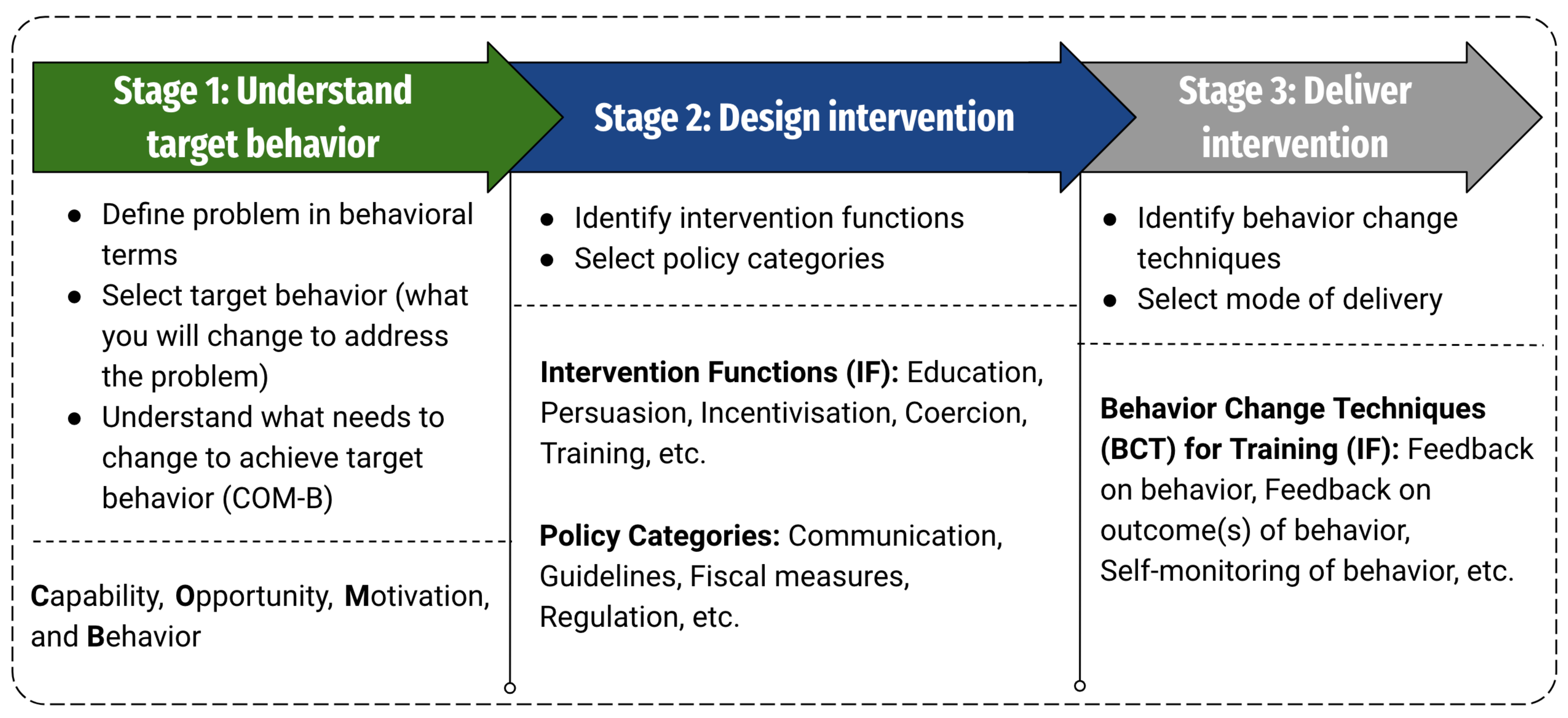}
	\caption{The Three Stages of the Behavior Change Wheel including some intervention functions, policy categories, and behavior change techniques, as indicated  in Fig.~\ref{fig:bcw}.}
	\label{fig:bcwstages}
\end{figure*}

Organizations interested in promoting software security practices across different teams and stakeholders should carefully consider developers' behaviors and identify their perceptions and attitudes regarding their capabilities, opportunities available to them, and their motivations for adopting software security practices. Based on our findings, we propose \frameworkname, a framework that consists of a comprehensive list of \numberofdrivers drivers that represent what needs to change or happen so that the adoption of software security practices occurs. In Table~\ref{tab:drivers}, we present the complete list of drivers. Following the Behavior Change Wheel (BCW) approach, these drivers are the starting point to design interventions to foster an effective organizational security culture and motivate developers to write secure code. As we described in Section~\ref{sec:background}, we recommend that organizations use \frameworkname to conduct the steps indicated in Figure~\ref{fig:bcwstages}.

	\cooltitle{Stage 1: }The goal of this stage is to use the \frameworkname framework to identify which drivers apply to the organization's particular context. To achieve this, organizations should conduct focus groups or semi-structured interviews, considering the perspectives of most stakeholders, from developers to managers. In the case of big corporations, where collecting data could be challenging due to the significant number of stakeholders involved in ensuring a secure product, another potential mechanism is to conduct a survey. A survey could help collect the level of agreement or disagreement concerning the drivers influencing stakeholders to adopt software security practices. In the end, the output of this stage is a subset of drivers that are relatable to the organization's context. 
	
	\cooltitle{Stage 2: }The goal of this stage is to identify the most appropriate intervention functions and then select their respective policy categories, which will serve as the foundation to design and deliver an intervention. 
	
	To select the most relevant intervention functions, organizations should use the subset of drivers identified in \textit{Stage 1} to shift from diagnosis to intervention. Organizations need to build a matrix using the COM-B model components and the 9 intervention functions proposed by the BCW. As detailed in Section~\ref{sec:background}, the COM-B model components to take into account are \textit{Technical Capabilities, Non-technical Capabilities, Psychological Capabilities, Technical Opportunities, Social Opportunities, Reflective Motivations, and Automatic Motivations}, and the intervention functions to consider are \textit{Education, Persuasion, Incentivisation, Coercion, Training, Restriction, Environmental restructuring, Modeling, and Enablement}. By using this matrix, organizations should be able to identify the intervention functions associated with their context. For example, our study identified a strong relationship between \textit{Training} and \textit{Psychological capabilities}. Developers perceived that having tailored security training is crucial for building confidence in implementing software security practices.
	
	The next step in developing an intervention strategy is to select the policy categories that can help deliver and implement the intervention. The BCW indicates which policy categories are the most suitable in supporting each intervention function. Following these indications, organizations should select the most appropriate policy categories associated with their intervention functions. For example, concerning \textit{Training}, some essential policy categories are \textit{Guidelines, Fiscal measures, Regulation, Legislation, and Service provision}. A complete list of the 7 policy categories and their definitions is detailed in Table~\ref{tab:policycategory}.      
	
	\cooltitle{Stage 3: }The goal of this stage is to determine which Behavior Change Techniques (BCTs) can support the delivery of the identified intervention functions under the relevant policy categories. Michie et al.~\cite{michie2014} defines a BCT as an active component of an intervention designed to change behavior. BCTs are also observable, replicable, and irreducible components of an intervention. Michie et al.~\cite{michie2013} introduces a taxonomy of 93 BCTs as a method for specifying interventions. With this knowledge, organizations may select the most frequently used BCTs for their particular intervention functions. For example, in the case of the \textit{Training} intervention function, some of the most frequently used BCTs are \textit{Feedback on behavior, Feedback on outcome(s) of behavior, Monitoring of behavior by others without evidence of feedback, Monitoring of outcome of behavior by others without evidence of feedback}, and \textit{Self-monitoring of behavior.} For example, a description of the content of an intervention would be ``Organizations should set the goal of \textit{training developers} to identify potential risks in their source code, \textit{enabling} them to understand security standards, associate those standards with threats and compliant code examples.''

\subsubsection{What organizations need to take into account when understanding developers' perspectives}

Our work revealed several attitudes and beliefs that are useful for organizations and researchers to understand the developer mindset behind software security practices. First, developers often perceive the \textit{disadvantages} of adopting security more quickly than the immediate benefits, e.g., time-consuming, delay feature delivery, add restrictions to their workflow, etc. Developers' mindsets typically have the \textit{time-pressure} concern as their goals focus on delivering features. If security is not baked into the development pipeline, it will usually be \textit{overlooked} if there are competing priorities. Additionally, developers are influenced to adopt security practices when other team members embrace the same practices. However, \textit{management} becomes a \textit{deterrent} if they do not support developers' proactiveness towards security. When a security role is present on the team, developers usually \textit{ignore} or \textit{delegate} security practices as they consider those practices outside their core responsibilities.

Furthermore, developers might self-assess their security knowledge and capabilities with high scores. Extensive development experience can produce \textit{false confidence} in the ability to effectively conduct a security practice without knowing the necessary security skill set required. Finally, most developers who are passionate about security have experienced security exploits in the past. The \textit{fear} of suffering another attack is their primary motivation to keep their security knowledge up-to-date and advocate security concerns among peers. External events that impact \textit{developers' emotions} have a high likelihood of sticking in a developer's mind for a longer period of time and will shape their attitude towards understanding and evaluating future risks. By default, it is relevant to acknowledge that it is not part of the developer's mindset to \textit{think upfront about risks} and what could go wrong when customers misuse the applications they develop. Therefore, it is essential that any corporate security training highlight these potential scenarios. 

\subsubsection{What organizations need to take into account when understanding security specialists' perspectives}

Developers and security professionals have different mindsets. Security specialists are usually more \textit{pessimistic} because they tend to prioritize risks over product features. Moreover, since security is the day-to-day work for a security professional, they usually employ \textit{technical communication} that developers typically are not familiar with. As a result, their perspectives concerning the impact of patching a security vulnerability might considerably differ from developers or engineers. Finally, since a security professional's primary focus is on identifying potential risks/threats and complying with security guidelines, their biased perspective might make it more challenging to see the overall impact of any change on the application. This situation becomes a threat in the organization when security teams analyze the trade-offs of applying security without the collaboration of developers and software architects.  

\subsubsection{Additional considerations organizations should be aware of when applying the framework}

The following considerations were emphasized by our participants during the member-checking sessions. 

\cooltitle{A security diagnosis should be conducted periodically.} Using the COM-B model as a diagnosis tool implies identifying what drives developers to adopt software security practices. However, practitioners' perceptions, attitudes, and motivations can change over time due to external factors or events. In this scenario, the applicability of specific drivers to the current organization's context could be outdated. Therefore, it is essential to acknowledge these potential changes and periodically monitor them to ensure that the strategies or interventions are not based on inaccurate facts or observations.  

\cooltitle{A \textit{just enough security} approach for startup companies.} In the case of startup companies, the concern around adopting software security practices when developing a product-market fit is only addressed if the targeted consumer demands it. Otherwise, other non-functional requirements will have higher priority. Security specialists recommend startups with limited resources focus on delivering a \textit{just enough} secure product, and if required, invest in training developers in security practices instead of hiring specialized resources. If available, external incentives, such as IRAP \footnote{https://nrc.canada.ca/en/support-technology-innovation/about-nrcindustrial-research-assistance-program} in Canada, may be beneficial to introduce security testing practices provided by external parties to guarantee a certain level of quality in terms of security. This opportunity will also help developers acknowledge their typical security mistakes and which topics to focus on in their self-directed security learning process.

\cooltitle{Non-technical skills matter.} Practitioners highlighted how it is essential to be aware of the security skill set required to implement security in the software development pipeline. Although most of our participants' first thoughts pointed out several security technical skills, they also mentioned different challenges related to interpersonal skills that, surprisingly, are usually not discussed in organizational security training. And when overlooked, it could seriously impact how security is handled within the organization, especially when multiple stakeholders are involved in security decisions and product development. It is worth mentioning that most security challenges are indirectly related to conflict management, negotiation skills, communication ability, and empathy. Therefore, we encourage organizations to include topics related to non-technical skills as part of any security training program, which is helpful in the context of security and boosts collaboration and productivity in the company. 

\subsection{Recommendations for Developers When Adopting Security Practices}

In addition to organizational recommendations, the drivers identified in our study point to specific recommendations for developers:  

\begin{itemize}
	\item Passionate software developers care about the quality of their code. Security is an essential quality aspect of any software product, and its adoption gradually grows with experience. In addition, when learning a new software technology, e.g., a programming language or a framework, developers should take the necessary time to learn how to use it securely, not just assume that default configurations or standard ways to use it are secure.
	\item If security is not part of the organizational culture, this can be an excellent opportunity to advocate the relevance of baking security into the development workflow. Developers should look for sponsorship at the management level. With management's support, any effort towards security will be more straightforward and significant. 
	\item When advocating for security, there can be some reluctance to change. Different stakeholders have different perspectives and priorities regarding security. To advocate for security more effectively, developers need to acquire the ability to translate security threats into technical and business risks. Arguments based on risks are more compelling and easier to understand.
	\item The adoption of security practices has an undeniable social connotation. It is recommended that developers start advocating for security among their peers. Raising a collective need or concern for security from engineering teams will significantly impact management levels more than any individual approach.
	\item Including security in the software development pipeline introduces several policies and restrictions around tools, frameworks, and open-source components. Developers should recognize that having full autonomy in technical decisions will cause a chaotic development environment, making it challenging for any stakeholder to ensure a secure development workflow and product.    
	\item Developers should leverage their current working environment to customize their learning path. The presence of a security-specific role in the organization willing to mentor through peer-to-peer learning or osmosis is a practical and valuable learning approach. For example, some realistic scenarios are developers shadowing a security specialist or being indirectly involved in fixing security vulnerabilities.  
	\item Developers should take advantage of software security practices conducted within the organization. They are an important opportunity for developers to follow a learning-by-doing approach, master software security skills, and receive actionable feedback to improve their development practices, e.g., through security code reviews. 
	\item We recommend developers participate in communities of practice led by software security professionals. They are an active community in the industry, continuously organizing meet-ups, conferences, and workshops where they share valuable security resources and their experiences dealing with security vulnerabilities.   
	\item We recommend developers use software security practices such as code reviews to self-assess their software security skills. Feedback from security code reviews can help developers identify flaws in their coding practices and suggest topics to include in an organizational security training program.    
	\item Confidence in performing software security tasks is an essential aspect that developers achieve through proper training and experience. However, developers should be aware that overconfidence is a deterrent to improving their security practices, assimilating feedback, and maintaining the knowledge required to deal with security exploits that are getting more sophisticated and harmful.       
\end{itemize} 

\subsection{Recommendations for Security Specialists When Advocating for Security}

The drivers we identified in our study also suggest a number of recommendations for security specialists: 

\begin{itemize}
	\item The developers in our study perceived that most security guidelines are abstract and not developer friendly. We suggest security specialists facilitate the comprehension of security guidelines by highlighting the relationship between security threats, non-compliant code examples, and compliant solutions. In other words, when designing security guidelines, security professionals should use a technical vocabulary that developers are more familiar with.     
	
	\item Most developers typically access Q\&A forums, e.g., Stack Overflow\footnote{https://stackoverflow.com/}, to gather knowledge, share expertise, and address security and development concerns. Therefore, security professionals should proactively approach Q\&A forums to get one step closer to developers' communities, advocating for security while interacting with developers by building awareness of security tools, potential security threats, and the risks of overlooking secure coding practices.
	
	\item Developers demand tailored security training that reflects their particular information needs and software development context. Developers perceive generic security training as too abstract, time-consuming, and ineffective. When designing security training to meet developers' needs, security teams or security specialists should identify developers' typical security mistakes through their interactions with engineering teams. For example, the interactions during security code reviews can be a valuable source of information.
	
	\item Developers who are passionate about security are usually self-taught software security practitioners. However, the amount of resources and topics to learn is typically overwhelming. Therefore, security teams should facilitate developers' learning process by allowing them to know the rationale behind the compliance with security guidelines and the necessary skills to tackle their typical security mistakes. In this way, developers will learn faster, remember security guidelines easier, gradually build confidence in their technical skills, and reduce the chance of repeating the same security mistakes.  
	
\end{itemize}   	

\subsection{The Power of Behavioral Theories for Software Security Researchers}

Our work is related to developer-centric security research. Since software development is intensely human driven, researchers must consider developers' behaviors, attitudes, beliefs, perceptions, and motivations to introduce any positive change in software engineering practices. Our study highlights the opportunity for the software engineering research community to analyze software security challenges through the lens of behavioral science theories. Most efforts from industry and academia in the area of software security has focused on providing exceptional security standards, sophisticated security tools, free online learning resources, understanding the effectiveness of having specific security roles in the engineering teams, and encouraging shifting security to the left or earlier stages of the software development pipeline. However, little attention has been dedicated to allowing organizations to design effective interventions to foster the adoption of software security practices among developers. 

Additionally, our work exposes a holistic view of all behavioral aspects that affect how developers adopt a target behavior, such as \textit{capabilities}, \textit{opportunities}, and \textit{motivations}. Developers will be willing to adopt a new practice if they feel confident and capable of doing it, have the right opportunities and conditions in their work environment, and feel motivated to perform the target behavior. Our study opens the door to further research to introduce behavior change techniques to understand developer and other stakeholder behaviors. This knowledge will serve as the foundation to design interventions to motivate them to adopt software security practices.

\vspace{5px}

	\section{Threats to Validity}
	\label{sec:threatstovalidity}

In the following, we address the validity of this study in the context of qualitative research~\cite{guba1981, korstjens2018}.

\subsection{Transferability}
Transferability is the degree to which we can transfer our results to other contexts. Our study was based on semi-structured interviews gathering the experiences of \numberofinterviews software engineers and security specialists. Given that their experiences, companies, technology stacks, and business domains varied considerably, the drivers identified in our study should fit most software development organizations. However, we did not include open-source developers in our work, geographical and cultural determinants to analyze our data, such as cultural differences in power distance and individualism-collectivism~\cite{taras2010}. In addition, we did not consider group dynamics that might influence security adoption, such as organization identity, organization climate, and culture~\cite{schneider2012}. Therefore, we suggest future research to understand whether the open-source community considers our findings relevant for their context and how geographical, cultural, and group dynamics-related factors may influence developers' adoption of security practices.

\subsection{Credibility}
Credibility concerns whether the research findings are correctly drawn from the original data. We applied three strategies to ensure credibility: (a) the list of drivers was iteratively developed by two researchers and examined by an expert reviewer in each iteration. In addition, two researchers performed the open coding of the transcribed interview data. At the beginning of the open coding process, two open coding iterations were conducted to align the perspectives of both researchers. After achieving at least 75\% inter-rater agreement, both researchers started coding independently, then (b) the set of drivers, categories, and findings were discussed several times between all the authors of this paper to mitigate bias from any particular researcher involved in the study, and (c) the drivers were validated through 12  member-checking sessions.  

\subsection{Confirmability}
Confirmability is the degree to which other researchers can confirm the findings. We do not have the participants' permission to share the transcriptions of the interviews. However, we tried to show as much evidence as possible for each driver by quoting participants when describing our results. Our interview script is available in our online appendix~\cite{appendix}. 
	\section{Conclusions}
	\label{sec:conclusions}

Adopting software security practices is a significant concern for any industry. The exponential increase in security vulnerabilities exploited by malicious hackers pushes the need to understand why software security is neglected and why developers persist in introducing security flaws into their applications. In this study, using the lens of the COM-B model, we systematically explore what needs to change or happen so the adoption of software security practices occurs. As a result, we propose \frameworkname, which consists of a comprehensive set of \numberofdrivers drivers that describes the software security adoption phenomena. Our work is the first to introduce a behavioral change approach to understand developers' behaviors when adopting software security practices. Using \frameworkname as a starting point, we foresee that organizations will be able to design appropriately geared interventions following the Behavioral Change Wheel framework. We hope our study insights will help organizations help developers write better, more secure code, ensuring a reliable and secure software product.

\vspace{5px}


%


\ifCLASSOPTIONcompsoc
  \section*{Acknowledgments}
\else
  \section*{Acknowledgment}
\fi

The authors would like to thank the 28 interviewees for their availability in this study, Dr. James Gibson for his remarkable insights in behavioral psychology at the beginning of our study, and the members of the CHISEL group at UVic for their invaluable feedback. We also acknowledge the support of the Natural Sciences and Engineering Research Council of Canada (NSERC).




\end{document}